\documentclass[]{aastex63}
\pdfoutput=1

\received{2019/08/30}
\revised{2019/10/14}
\accepted{2019/10/19}
\submitjournal{ApJ}

\shorttitle{dust reverberation of quasars}
\shortauthors{Minezaki et al.}

\begin{document}

\title{REVERBERATION MEASUREMENTS OF THE INNER RADII OF THE DUST TORI IN QUASARS}

\correspondingauthor{Takeo Minezaki}
\email{minezaki@ioa.s.u-tokyo.ac.jp}

\author{Takeo Minezaki}
\affiliation{Institute of Astronomy, School of Science, University of Tokyo \\
2-21-1 Osawa, Mitaka, Tokyo 181-0015, Japan}

\author{Yuzuru Yoshii}
\affiliation{Institute of Astronomy, School of Science, University of Tokyo \\
2-21-1 Osawa, Mitaka, Tokyo 181-0015, Japan; PI of the MAGNUM project}
\affiliation{Steward Observatory, University of Arizona \\
933 North Cherry Avenue, Room N204, Tucson, AZ 85721-0065, USA}

\author{Yukiyasu Kobayashi}
\affiliation{National Astronomical Observatory of Japan\\
2-21-1 Osawa, Mitaka, Tokyo 181-0015, Japan}

\author{Shota Sugawara}
\affiliation{Institute of Astronomy, School of Science, University of Tokyo \\
2-21-1 Osawa, Mitaka, Tokyo 181-0015, Japan}
\affiliation{Department of Astronomy, School of Science, University of Tokyo \\
7-3-1 Hongo, Bunkyo-ku, Tokyo 113-0013, Japan}

\author{Yu Sakata}
\affiliation{Institute of Astronomy, School of Science, University of Tokyo \\
2-21-1 Osawa, Mitaka, Tokyo 181-0015, Japan}
\affiliation{Department of Astronomy, School of Science, University of Tokyo \\
7-3-1 Hongo, Bunkyo-ku, Tokyo 113-0013, Japan}

\author{Keigo Enya}
\affiliation{Institute of Space and Astronautical Science, Japan Aerospace Exploration Agency \\
3-1-1, Yoshinodai, Sagamihara, Kanagawa 229-8510, Japan}

\author{Shintaro Koshida}
\affiliation{Subaru Telescope, National Astronomical Observatory of Japan\\
650 North Aohoku Place, Hilo, HI 96720, USA}

\author{Hiroyuki Tomita}
\affiliation{Institute of Astronomy, School of Science, University of Tokyo \\
2-21-1 Osawa, Mitaka, Tokyo 181-0015, Japan}

\author{Masahiro Suganuma}
\affiliation{Institute of Space and Astronautical Science, Japan Aerospace Exploration Agency \\
3-1-1, Yoshinodai, Sagamihara, Kanagawa 229-8510, Japan}

\author{Tsutomu Aoki}
\affiliation{Kiso Observatory, Institute of Astronomy, School of Science, University of Tokyo \\
10762-30 Mitake, Kiso, Nagano 397-0101, Japan}

\author{Bruce A. Peterson}
\affiliation{Mount Stromlo Observatory, Research School of Astronomy and Astrophysics, Australian National University \\
Weston Creek P.O., ACT 2611, Australia}

\begin{abstract}
 We present the results of a dust-reverberation survey of
 quasars at redshifts $z<0.6$.
 We found a delayed response of the $K$-band flux variation
 after the optical flux variation in 25 out of 31 targets,
 and obtained the lag time between them for 22 targets.
 Combined with the results for nearby Seyfert galaxies,
 we provide the largest homogeneous collection of
 $K$-band dust-reverberation data
 for 36 type 1 active galactic nuclei (AGNs).
 This doubles the sample and includes
 the most distant AGN and the largest lag so far measured.
 We estimated the optical luminosity of the AGN component
 of each target using three different methods:
 spectral decomposition,
 the flux-variation-gradient method,
 and image decomposition.
 We found a strong correlation between
 the reverberation radius for the innermost dust torus
 and the optical luminosity over a range of
 approximately four orders of magnitude in luminosity,
 as is already known for Seyfert galaxies.
 We estimated the luminosity distances of the AGNs
 based on their dust-reverberation lags,
 and found that the data in the redshift--distance diagram 
 are consistent with the current standard estimates
 of the cosmological parameters.
 We also present the radius--luminosity relations
 for isotropic luminosity indicators such as
 the hard X-ray (14--195 keV), [O{\small IV}] 25.89 $\mu$m,
 and mid-infrared (12 $\mu$m) continuum luminosities,
 which are applicable to obscured AGNs.
 \end{abstract}


\section{Introduction} \label{sec:intro}

A dust torus surrounding a supermassive black hole,
an accretion disk, and a broad emission-line region (BLR)
is an important structure in an active galactic nucleus (AGN).
The dust torus is a potential gas reservoir
for fueling mass to the accretion disk,
where the enormous radiation energy of the AGN is produced.
It is also a key structure
in the unified scheme of AGNs,
in which the diversity of type 1 and type 2 AGNs is explained by
the viewing angle and obscuration by the dust torus
\citep[e.g.,][]{1993ARA&A..31..473A}.
Dust in the torus is illuminated by the strong UV--optical continuum
emission from the accretion disk, and it absorbs the radiation energy
and then re-radiates it as thermal emission at infrared wavelengths.
Since the dust temperature is limited by sublimation of the dust
\citep[at approximately $1700$--$2000$ K for graphite grains:][]{1977AdPhy..26..129H,1977ARA&A..15..267S,2018MNRAS.474.1970B},
the hottest dust is thought to be located in the innermost dust torus,
producing near-infrared thermal emission,
and the size of the hot-dust region is expected
to be correlated with the accretion-disk luminosity
as $r\propto L^{0.5}$, according to the radiation equilibrium of dust
\citep{1987ApJ...320..537B}.

Dust-reverberation mapping of an AGN
is an important tool for investigating the spatially unresolved
structure of the innermost dust torus.
It measures the time lag between a flux variation of the UV--optical continuum
emission from the accretion disk and that of the near-infrared thermal emission
from the innermost dust torus,
which is interpreted as the light-travel time from the accretion disk
to the dust torus.
The radius of the innermost dust torus
--- or its more detailed geometrical structure ---
can be inferred from the time lag
or the delayed response of the dust emission.
Pioneering work has been performed on this topic since the 1970s
\citep{1971MNRAS.153...29P,1980Natur.284..410L,1989ApJ...337..236C,1992A&A...256..375B,1992MNRAS.256P..23G,1993AstL...19..416O,1993ApJ...409..139S,1999AstL...25..483O,1996ApJ...465L..87N,2004MNRAS.350.1049G}.

A systematic dust-reverberation survey for a number of type 1 AGNs
has been performed
by the Multicolor Active Galactic Nuclei (MAGNUM) project
\citep{2002ntto.conf..235Y,2003AAS...202.3803Y}.
It established observationally a radius--luminosity relation
for the innermost dust torus based on 49 dust-reverberation lags
observed in the $K$ band ($\lambda = 2.2\ \mu$m)
for 17 nearby Seyfert galaxies
\citep{2004ApJ...600L..35M,2004ApJ...612L.113S,2006ApJ...639...46S,2014ApJ...788..159K},
demonstrated the accretion-disk component in the near-infrared emission
and obtained the lag between the flux variations
of the dust emission at different wavelengths
\citep{2006ApJ...643L...5M,2006ApJ...652L..13T},
and examined the long-term variation of the dust-reverberation lag
for an AGN \citep{2009ApJ...700L.109K,2014ApJ...788..159K}.
Recent dust-reverberation studies
have also been conducted to make a progress in studying
the geometry and structure around the central region of the dust torus
\citep{2008A&A...486..411S,2011MNRAS.415.1290L,2015MNRAS.454..368L,2013A&A...557L..13S,2015A&A...578A..57S,2014A&A...561L...8P,2015A&A...576A..73P,2015ApJ...814L..12J,2015ApJ...801..127V,2017MNRAS.467.1496O,2019AstL...45..197O,2018MNRAS.475.5330M,2018A&A...620A.137R,2019arXiv190408946K,2019MNRAS.485.2573K,2019MNRAS.489.1572L},
and \citet{2019arXiv190911101L} very recently presented
the results of a systematic dust-reverberation survey
at longer wavelengths ($3.4$ and $4.5\ \mu$m)
for the Palomar--Green (PG) quasars \citep{1983ApJ...269..352S}.

In addition,
a possible application of dust reverberation
to the distance measurements of AGNs
was proposed at the end of the 20th century
to investigate the cosmic expansion
based on the radius--luminosity relation
\citep{1998SPIE.3352..120K,1999OAP....12...99O,2001ASPC..224..149O,2002ntto.conf..235Y}.
The MAGNUM project was started in order to establish
this dust-reverberation--distance method
and thereby constrain the cosmological parameters.
After the collection of a homogeneous set of dust-reverberation data
for 17 Seyfert galaxies by the MAGNUM project,
\citet{2014ApJ...784L..11Y}
built a model for the radiation equilibrium of dust grains
in the innermost dust torus to obtain the luminosity distances of AGNs
without requiring any distance ladder.
They estimated the Hubble constant from these data
and found it to be in good agreement with the current standard estimates.
\citet{2017ApJ...842L..13K} compared
the distance calibration of \citet{2014ApJ...784L..11Y}
to the distances of Type Ia supernovae that occurred in the AGN host galaxies
and found that they are consistent.
These results indicate that the distance indicator based on
dust reverberation is a promising new tool for investigating
the cosmic expansion.
Now, large new dust-reverberation surveys have been started or been proposed
\citep{2014ApJ...784L...4H,2017MNRAS.464.1693H}.

In this study,
we present the results of $K$-band
dust-reverberation observations for 31 quasars at redshifts $z<0.6$
performed by the MAGNUM project.
We found delayed responses of the near-infrared flux variations
for 25 targets and obtained dust-reverberation lags for 22 of them.
They include the most distant AGN with a dust-reverberation lag
and a luminous quasar with the largest lag measured so far in the $K$ band.
We combined these data with the results for nearby Seyfert galaxies
\citep{2014ApJ...788..159K}
to examine the radius--luminosity relation
for the innermost dust tori for Seyfert galaxies and quasars.
We also estimated their luminosity distances based on
the dust-reverberation lags
in order to constrain the cosmic expansion.
In Section 2, we describe the target AGNs and 
the procedures of observation, reduction, and photometry,
and present the optical and near-infrared light curves.
In Section 3, we measure the lag time between the optical and
near-infrared flux variations.
In Section 4, we estimate the optical luminosity of the AGN component
and examine the radius--luminosity relation for the innermost dust torus.
In Section 5, we calculate the dust-reverberation distance
for the targets at $z<0.6$ and fit the diagram of redshift
versus luminosity distance to the cosmological models.
We summarize the results in Section 6.
We assume a cosmology with $H_0=73$ km s$^{-1}$ Mpc$^{-1}$,
$\Omega_0=0.27$, and $\lambda_0=0.73$ 
\citep{2007ApJS..170..377S}
throughout this paper, except in Section 5.

 \section{Observations and Photometry} \label{sec:obs}
 \subsection{Targets} \label{subsec:target}
 In this paper, we report the results for $31$ type 1 AGNs
 monitored by the MAGNUM project.
 They include not only quasars but also less luminous type 1 AGNs.
 We limited the redshift to $z<0.6$ so that
 the thermal emission from the dust torus would be significant
 when it is observed in the near-infrared $K$ band.
 The targets and their basic parameters are listed in Table \ref{tab:objdata}.
 Based on target visibility
 and the presence of moderately bright stars
 within the field of view of the camera
 to serve as photometric references,
 we selected $28$ type 1 AGNs from
 the Early Data Release (EDR) catalog
 of the Sloan Digital Sky Survey \citep[SDSS, ][]{2002AJ....123..485S},
 the 2-degree Field (2dF) QSO Redshift Survey
 \citep[2QZ,][]{2001MNRAS.322L..29C},
 the Large Bright Quasar Survey
 \citep[LBQS,][]{1995AJ....109.1498H},
 and
 the ninth edition of the Catalog of Quasars and Active Nuclei
 by Veron-Cetty and Veron \citep[VV9,][]{2000cqan.book.....V}
 \footnote{No radio-loud quasar was selected
 from the VV9 catalog because
 it is possible for a nonthermal variable flux component
 in the near-infrared to be present in addition to
 the thermal emission from the dust torus
 in radio-loud quasars
 \citep[e.g.,][]{1999MNRAS.310..571M,2007MNRAS.375.1521M,
 2002ApJS..141...45E}.}.
 In addition, we selected three of the PG quasars,
 for which the broad emission-line (BEL) lag
 has been measured by reverberation-mapping observations
 \citep{2000ApJ...533..631K}.

 \subsection{Observations} \label{subsec:obs}
 We conducted monitoring observations using
 a multicolor imaging photometer (MIP) mounted
 on the MAGNUM 2 m telescope at the Haleakala Observatories in Hawaii
 \citep{1998SPIE.3354..769K}.
 The MIP is capable of simultaneous imaging in one optical band
 (selected from among $U$, $B$, $V$, $R$, and $I$)
 and one near-infrared band (selected from among $Y$, $J$, $H$, and $K$)
 by splitting the incident beam into two different detectors.
 We employed an SITe CCD ($1024\times 1024$ pixels) 
 and an SBRC InSb array ($256\times 256$ pixels),
 respectively, as the optical and infrared detectors.
 The MIP field of view was $1.5\times 1.5$ arcmin$^2$.

 In order to measure the dust-reverberation lag,
 we monitored the targets most frequently
 in the $K$ band and one optical band.
 We selected the latter from among the $V$, $R$, and $I$ bands,
 depending on the target redshift,
 so that the rest-frame wavelength
 of the observing band was close to the $V$ band.
 We monitored the targets less frequently
 in the next bluer optical bands that we used to decompose
 the AGN and host-galaxy fluxes (see Section \ref{subsec:host}).
 We started regular monitoring observations in 2001--2003
 and finished them in 2007.
 We planned to monitor the less luminous targets more frequently,
 because both the timescale of the flux variation
 and the dust-reverberation lag
 are expected to be shorter for such targets.
 The observational parameters for these sources
 are also listed in Table \ref{tab:objdata}.

 \subsection{Reduction and Photometry} \label{subsec:phot}
 We reduced the images using IRAF.\footnote{
 IRAF is distributed by the National Optical Astronomy Observatories,
 which are operated by the Association of Universities for Research
 in Astronomy, Inc., under cooperative agreement with the National
 Science Foundation.}
 We followed the standard procedures for image reduction,
 such as correction for bias or sky subtraction and flat-fielding.

 For the 28 targets selected from the SDSS EDR, the 2QZ, the LBQS,
 and the VV9, we observed the target AGN and the nearby field stars
 within the field of view simultaneously.
 We then measured the target flux with respect to the field stars 
 using aperture photometry with a circular aperture of $\phi = 4.15$ arcsec
 diameter and a sky reference area of $\phi= 11.1$--$13.9$ arcsec annulus.

 On the other hand, we found no suitable stars to serve as
 photometric references within the fields of view around the target PG quasars.
 We therefore observed the target and reference stars
 outside the field of view alternately,
 and we measured the target flux
 with respect to those reference stars,
 as has been done for nearby Seyfert galaxies
 \citep{2014ApJ...788..159K}.
 We used a larger circular aperture of $\phi = 8.3$ arcsec diameter
 for those targets, because the target and the reference star were
 not observed simultaneously, and the point-spread functions (PSFs)
 for their images were possibly slightly different.

 We calibrated the fluxes for the reference stars with respect to
 the photometric standard stars of
 \citet{1992AJ....104..340L} and \citet{1998AJ....115.2594H}
 for the $V$ and $K$ bands, respectively.
 Finally, we corrected for Galactic extinction
 using the NASA/IPAC Extragalactic Database
\citep[NED; based on][]{2011ApJ...737..103S}.
 The light curves of the 31 target AGNs are presented
 in Figure \ref{fig:lc1},
 and the corresponding data are listed in Table \ref{tab:lcdata}.

 \section{dust-reverberation analysis} \label{sec:lag}
 As shown in Figure \ref{fig:lc1},
 most targets showed a delayed response of the $K$-band flux variation
 after the optical flux variation,
 which indicates the reverberation of the thermal emission
 from the dust tori in quasars.
 We estimate the lag time between the optical and $K$-band flux variations
 as described below for the targets
 that showed apparent delayed $K$-band responses.
 In this study, we estimated the lag time for each target
 during the whole monitoring period
 and did not examine the time variation of the dust-reverberation lag
 in a target,
 because many luminous targets showed relatively slow flux variations.

 \subsection{Accretion-disk Component in the K-band Fluxes} \label{subsec:subad}
 In addition to the thermal emission from the dust torus,
 the continuum emission from the accretion disk is thought to
 contribute to the near-infrared flux of a type 1 AGN.
 Contamination by the accretion-disk component
 makes the lag between the optical and near-infrared flux variations 
 more or less shorter than the actual lag of the dust-torus emission response
 \cite[e.g.,][]{2006ApJ...643L...5M, 2006ApJ...652L..13T, 2011MNRAS.415.1290L,2014ApJ...788..159K},
 because the near-infrared flux variation from the accretion disk
 is almost synchronous with its optical flux variation.
 Since the contribution from the accretion-disk component
 becomes larger at shorter wavelengths,
 possible bias in the measurement of the dust-reverberation lag
 is more serious for the quasars at larger redshifts.

 We assumed that the continuum emission of the accretion disk
 has a power-law spectrum and the flux changes simultaneously
 at all wavelengths. Then, the near-infrared fluxes of
 the accretion-disk component can be estimated as
 \begin{equation}
  \label{eqn:ADcomp}
 f_{{\rm AD},K}(t)=\left(\frac{\nu_{K}}{\nu_{X}}\right)^{\alpha _{\rm OIR}} f_{{\rm AD},X}(t)\, ,
 \end{equation}
 where $X$ represents the optical observing band for each target.
 We subtracted $f_{{\rm AD},K}(t)$ from the $K$-band fluxes
 in the procedure for measuring the dust-reverberation lag,
 as described in the next subsection.

 We adopted the power-law index $\alpha _{\rm OIR}=+0.1$
 following \citet{2014ApJ...784L..11Y},
 based on the cross multiple-regression analysis
 of the $V$- and $K$-band flux variations of local Seyfert 1 galaxies
 \citep[$\alpha _{VK}= +0.1\pm 0.11$;][]{TomitaPHD,2006ApJ...652L..13T}.
 We also estimated the dust-reverberation lags
 assuming $\alpha_{\rm OIR}=0.0$ and $+1/3$,
 as had been done by \citet{2014ApJ...788..159K},
 to examine possible systematic differences depending on $\alpha_{\rm OIR}$.
 We note that $\alpha _{\nu} = +1/3$ is the power-law index
 of the standard accretion-disk model \citep{1973A&A....24..337S}
 in the long-wavelength limit.

 \subsection{Measuring the Dust-reverberation Lag} \label{subsec:jav}
 We measured the dust-reverberation lag
 using the JAVELIN software \citep{2011ApJ...735...80Z},
 which has been widely used in recent reverberation studies.
 It has the advantage that uncertainties in the interpolation
 of the light-curve data can be estimated self-consistently
 in a statistical model of the variability.
 It adopts a damped random-walk model for
 flux variations of the source emission;
 this is considered to be a good approximation
 for the flux variations of the UV--optical continuum emission
 of type 1 AGNs \citep[e.g.,][]{2009ApJ...698..895K,2010ApJ...708..927K,2010ApJ...721.1014M,2012ApJ...753..106M,2013ApJ...765..106Z}.
 It assumes a top-hat transfer function in the transfer
 equation of the reverberation model,
 and it fits the source and response light-curve data
 using the Markov Chain Monte Carlo method
 to obtain the likelihood distribution for the lag time
 as well as those for the other parameters of the reverberation model.
 Recent enhancements of the capability of photometric
 reverberation mapping for the broad emission lines \citep{2016ApJ...819..122Z}
 enabled us to fit the data for the $K$-band flux variation
 with the delayed response of the dust-torus emission
 superimposed with $f_{{\rm AD},K}(t)$,
 which is proportional to (and thus synchronous with)
 the optical flux variation
 \citep{2019arXiv190408946K}.
 We fixed the proportionality factor
 $f_{{\rm AD},K}(t)/f_{{\rm AD},X}(t)$
 as determined by Equation \ref{eqn:ADcomp}
 with $\alpha _{\rm OIR}=+0.1$.

 We determined the best estimate and the $\pm 1\sigma $ errors
 for the dust-reverberation lag
 from the median, 15.9th percentile, and 84.1th percentile
 of the likelihood distribution for the lag time
 that was located around the peak of
 the cross-correlation function (CCF)
 of the optical and near-infrared flux variations.
 When the multiple peaks in the likelihood distribution
 are found around the CCF peak,
 we calculated the best estimate of the lag
 by the averaging the lags of the 15.9th and 84.1th percentiles.
 We estimated the CCF by using the PYCCF software
 \citep{2018ApJ...857...86S}, which calculates
 the interpolation CCF
 \citep[ICCF; e.g.,][]{1998PASP..110..660P}.
 We subtracted the accretion-disk component
 that we estimated from the optical flux assuming $\alpha _{\rm OIR}=+0.1$
 from the $K$-band flux before calculating the CCF.

 \subsection{Results} \label{subsec:dustlag}
 Figure \ref{fig:lag1} shows the likelihood distributions of
 the dust-reverberation lag obtained by JAVELIN overlaid
 on the CCFs for the 25 targets for which we found delayed responses
 of the $K$-band flux variation.
 We succeeded in obtaining the lag times for 22 targets.
 They include the most distant quasar with a dust-reverberation lag
 measured so far ($z\sim 0.6$),
 and the lag times include the largest dust-reverberation lag in the $K$ band
 ever obtained ($\sim 1.6$ yr in the observer frame).
 Combined with the results on 17 Seyfert galaxies
 from \citet{2014ApJ...788..159K},
 we here provide the largest homogeneous collection
 of $K$-band dust-reverberation data for AGNs,
 which has more than double the number of targets.
 The lag times and their errors are listed in Table \ref{tab:lagdata}.

 In addition,
 we estimated a lower limit to the dust-reverberation lag
 and the uncertain lags for the remaining three targets
 with apparent delayed responses in the $K$ band.
 In the former case  (2QZ J1225$-$0101),
 the CCF showed a very broad plateau
 at the lag time of $\sim 200$--$900$ days,
 but the number of the photometric data points
 in the overlapping time span
 of the lag-shifted optical and $K$-band light curves
 became so small at large lags 
 that the significance of the cross-correlation may be suspect,
 even within the CCF plateau.
 Since it was difficult to select
 one of the multiple peaks in the likelihood distribution
 obtained by JAVELIN in the range of the CCF plateau,
 we estimated the 0.135 percentile of the likelihood distribution
 as the $3\sigma $ lower limit of the dust-reverberation lag,
 and listed this in Table \ref{tab:lagdata}.
 In the latter cases (PG 0844$+$349 and 2QZ J1013$+$0028),
 the CCFs showed bimodal distributions,
 and it was uncertain which peak was significant.
 For reference, the lag times and their errors estimated from
 the JAVELIN likelihood distributions associated with
 the shorter lag in the CCF are also listed in Table \ref{tab:lagdata}.

 Unfortunately, we found no delayed response
 in the $K$-band light curves for six targets.
 For three targets
 (LBQS 0023$+$0228, 2QZ J1220$-$0119, and 2QZ J1438$-$0116),
 the amplitudes of possible optical flux variations were small
 while errors in the $K$-band photometries were apparently large.
 It was therefore unclear whether there was any delayed response.
 For two targets (2QZ J1032$-$0233, LBQS 1339+0210),
 the $K$-band fluxes did not change gradually with time
 but were spread by more than the estimated photometric errors.
 This made us suspect unaccounted systematic errors
 in the $K$-band photometries,
 and may be responsible for the lack of apparent delayed response.
 For the last target (IRAS F21256+0219),
 the $K$-band fluxes were fairly stable with time
 while the amplitude of the flux variation in the optical bands
 seemed to be small and not well sampled,
 which would lead to no detection of a delayed response.
 From these considerations,
 we consider dust reverberation to be ubiquitous,
 not only for Seyfert galaxies but also for quasars.

 In the last part of this section,
 we examine possible uncertainties in the dust-reverberation lag
 due to the accretion-disk component in the $K$-band flux.
 Figure \ref{fig:lagalpha} shows a comparison of
 the estimated lag times
 assuming $\alpha_{\rm OIR}=0.0$ and $+1/3$.
 We found that
 the lag times assuming $\alpha_{\rm OIR}=0.0$
 tended to be larger than
 those assuming $\alpha_{\rm OIR}=+1/3$, as expected
 (contamination by the accretion-disk component
 in the $K$-band flux is larger for smaller $\alpha_{\rm OIR}$).
 However, the difference was small,
 and in many cases the results were almost identical.
 The average difference was approximately $0.03$ dex,
 which is consistent with a similar study that estimated
 the dust-reverberation lags for Seyfert galaxies
 \citep{2014ApJ...788..159K}.

 \subsection{Redshift Correction for the Dust-reverberation Lag} \label{subsec:zcorr}

 We corrected the observed lag times for possible systematic
 changes in the dust-reverberation lag depending on redshift
 in order to compare them to those of AGNs in the local universe.
 One obvious source is cosmological time dilation,
 for which the observed time lag is corrected
 by multiplying by a factor of $(1+z)^{-1}$.

 Another factor is the wavelength dependence of the dust-reverberation lag.
 Since we observed the flux variation of thermal dust emission
 in the $K$ band,
 the rest-frame wavelength of the observed dust emission
 becomes shorter at larger redshifts.
 A radiation-equilibrium model of a dust grain
 directly illuminated by the accretion disk
 predicts $r \propto T^{2.8}$,
 where $T$ is the temperature of the dust grain
 and $r$ is the distance from the accretion disk \citep{1987ApJ...320..537B}.
 Assuming that the near-infrared emission of the dust torus
 at wavelength $\lambda $ is dominated by
 the thermal emission from dust at temperature $T$
 that peaks at $\lambda $,
 the wavelength dependence of the dust-reverberation lag
 is predicted to vary as $\Delta t\propto \lambda ^{2.8}$
 \citep{1993ApJ...409..139S,2001ASPC..224..149O}.
 This leads to a correction factor of $(1+z)^{2.8}$,
 opposite to that for cosmological time dilation.
 
 However, much smaller differences
 in the dust-reverberation lag at different wavelengths
 have been reported for multi-wavelength dust reverberation.
 Significant, but small differences were found in some AGNs
 \citep{TomitaPHD,2006ApJ...652L..13T,2011MNRAS.415.1290L,2015ApJ...814L..12J,2015ApJ...801..127V,2019arXiv190911101L},\footnote{
 \citet{2019arXiv190911101L} reported the ratios
 of the dust-reverberation radii in $\lambda=2.2$, $3.4$, $4.5$ $\mu$m
 as approximately 0.6:1:1.2 by comparing their results with
 those in the $K$ band of \citet{2014ApJ...788..159K}.
 When the $K$-band radius--luminosity relation for the optical luminosity
 of Equation \ref{eqn:rLfree} in this paper
 is converted to that for the bolometric luminosity
 using Equation 14 of \citet{2019arXiv190911101L},
 the ratio of the dust-reverberation radii in $\lambda=2.2$ and $3.4$ $\mu$m
 is estimated at approximately 0.8:1
 by comparing it with their result for $\lambda=3.4$ $\mu$m.}
 and no significant difference was found in some other AGNs
 \citep{2015MNRAS.454..368L,2015A&A...576A..73P,2015A&A...578A..57S,2018A&A...620A.137R,2019AstL...45..197O}.
 In some studies,
 small differences between lag times for different near-infrared bands
 were found,
 although it may be difficult to attribute
 the observed differences in lag times conclusively to
 the wavelength dependence of the dust-reverberation lag,
 because the accretion-disk component had not been subtracted
 from the near-infrared fluxes
 \citep{1989ApJ...337..236C,1993ApJ...409..139S,2004MNRAS.350.1049G,2008A&A...486..411S,2014A&A...561L...8P,2015OAP....28..175O,2017MNRAS.467.1496O,2018MNRAS.475.5330M}.

 In this study, we followed \citet{2014ApJ...784L..11Y}
 and formulated the correction factor empirically
 by fitting a function of redshift
 to the result of a cross multiple-regression analysis
 \citep{2006ApJ...652L..13T}
 applied to the multi-wavelength flux variations for six nearby Seyfert galaxies
 obtained by the MAGNUM project \citep{TomitaPHD}.
 They found that on average the difference in lag times 
 between the $H$- and $K$-band flux variations of the dust-torus emission
 was approximately 0.3 times the $K$-band dust-reverberation lag.
 Instead of the linear interpolation adopted by \citet{2014ApJ...784L..11Y},
 we used the factor $(1+z)^{\gamma }$ for the correction
 and obtained $\gamma =1.18$.
 By combining this with the correction factor for cosmological time dilation,
 we obtained the total correction factor $(1+z)^{\gamma -1}=(1+z)^{0.18}$. 
 The redshift-corrected lag times are also listed in
 Table \ref{tab:lagdata}.

 \section{Radius--Luminosity Relation for the Dust Torus} \label{sec:rL}
 Here we present dust-reverberation lags for 22 luminous type 1 AGNs,
 including quasars.
 Combined with the results for local Seyfert galaxies from
 \citet{2014ApJ...788..159K}, the MAGNUM project thus provides
 the largest homogeneous collection of $K$-band dust-reverberation data,
 which have been analyzed carefully and uniformly to obtain
 precise estimates of the lag times.
 Using these results, we examined the correlation between
 the dust-reverberation radius and the optical luminosity
 over ranges extending to larger radii and larger luminosities
 than those presented in previous studies
 \citep{2006ApJ...639...46S,2014ApJ...788..159K}.
 We also use the new data to update the correlation with
 the isotropic luminosity indicators for AGNs,
 which will be useful for studies of the structure in obscured AGNs
 \citep[e.g.,][]{2015ApJ...802...98M}.
 
 \subsection{Decomposition into the AGN and Host-galaxy Fluxes} \label{subsec:host}
 The flux obtained by aperture photometry contains a significant contribution
 from the host-galaxy starlight. We subtracted this from the optical flux 
 averaged over the monitoring period
 to obtain the flux from the AGN accretion disk.
 We applied three methods to the targets
 for the decomposition of the optical flux:
 spectral decomposition,
 the flux-variation-gradient method,
 and image decomposition.
 In all three methods, we corrected properly for Galactic extinction.
 
 \subsubsection{Spectral Decomposition by QSFit} \label{subsubsec:QSFit}
 We applied a spectral decomposition method to
 the targets for which spectra were available
 in the tenth Data Release of the SDSS
 \citep[SDSS DR10;][]{2014ApJS..211...17A}
 by using the Quasar Spectral Fitting package (QSFit) software
 \citep{2017MNRAS.472.4051C}\footnote{
 http://qsfit.inaf.it/
 }.
 QSFit fits the AGN optical spectrum with a combination of
 an AGN power-law continuum, the host-galaxy spectrum,
 the Balmer continuum, and the emission lines that include
 the blended iron complex.
 It fixes the power-law index at $\alpha _{\nu }=-0.3$
 in the redshift range of our targets ($z<0.6$),
 and uses a simulated 5 Gyr old elliptical galaxy template
 from the SWIRE template library\footnote{
 http://www.iasf-milano.inaf.it/\~polletta/templates/swire\_templates.html
 } \citep{1998ApJ...509..103S,2007ApJ...663...81P}
 as the host-galaxy spectrum.
 We subtracted the AGN power-law continuum calculated by QSFit
 from the SDSS DR10 spectrum and then integrated the residual spectrum
 with the response function of the observing band
 to estimate the host-galaxy flux.
 We transferred the observational errors in the SDSS DR10 spectrum data
 and the QSFit fitting error in the scaling factor of the power-law continuum
 to the error of the host-galaxy flux.

 We took the fitted parameters from the catalog of spectral properties
 of type 1 AGNs \citep[the QSFit catalog;][]{2017MNRAS.472.4051C}\footnote{
 http://qsfit.inaf.it/cat\_1.24/
 }
 for 12 targets that had already been processed with QSFit,
 and we processed another five targets with QSFit ourselves.
 The estimated host-galaxy fluxes and their errors are listed
 in Table \ref{tab:hostflux}.
 We did not apply QSFit to the remaining 14 targets,
 either because no SDSS DR10 spectra were available
 or because the fraction of ``good'' mask flags in the SDSS DR10 spectra
 was so small that QSFit did not function.

 \subsubsection{Decomposition by Flux Variation Gradient} \label{subsubsec:FVG}
 The multi-epoch flux data in two different optical bands
 obtained on the same night are known to show a tight linear correlation
 when they are plotted on a diagram of the flux in one band
 versus the flux in the other.
 This indicates that the optical continuum flux consists of
 a variable AGN flux with a constant spectral shape
 and a non-variable flux component due to the host galaxy
 \citep[e.g.,][]{1981AcA....31..293C,1992MNRAS.257..659W,
 1997MNRAS.292..273W,2010ApJ...711..461S}.
 Indeed, \citet{2010ApJ...711..461S} estimated the host-galaxy flux
 using the image-decomposition method applied to
 high-angular-resolution images of local type 1 AGNs
 and demonstrated that
 the non-variable component of the host-galaxy flux
 with some contribution from narrow emission lines
 is approximately located on a fainter extension
 of the linear regression line from the multi-epoch flux data
 in the flux-to-flux diagram.
 
 The host-galaxy flux in the aperture can thus
 be estimated from the intersection of
 the linear regression line from the multi-epoch flux data
 and a straight line from the origin,
 the gradient of which corresponds to the color of the host galaxy,
 as illustrated by \citet{1992MNRAS.257..659W}.
 We applied this flux-variation-gradient (FVG) method
 to our data
 as a second method for estimating the host-galaxy fluxes
 in the targets.
 We used the SWIRE 5 Gyr old elliptical template at the target redshift,
 which is adopted in QSFit,
 to calculate the flux ratio between
 the two different bands for the host-galaxy starlight,
 which is the gradient of the straight line from the origin.
 We estimated the error from the intersection of
 the straight line from the origin
 and the $\pm 1\sigma $ envelope of the linear regression
 of the multi-epoch flux data.
 
 We obtained the host-galaxy fluxes for 25 targets
 by using the FVG method.
 The estimated host-galaxy fluxes and their errors are listed
 in Table \ref{tab:hostflux}.
 We did not obtain the host-galaxy fluxes for the remaining six targets
 either because insufficient data points of multi-epoch flux data
 had been obtained simultaneously in the two different bands
 or because the flux variations in the optical band had small amplitudes.

 \subsubsection{Image Decomposition Using Hubble Space Telescope (HST) Images} \label{subsubsec:HST}
 We applied an image-decomposition method to four targets
 for which high-angular-resolution images were available
 in the Hubble Space Telescope (HST) archive.
 The PG quasars among our targets were observed
 with the High Resolution Channel of the Advanced Camera for Surveys (ACS/HRC)
 in the F550M band (Proposal IDs 9851 and 10833, PI Bradley Peterson),
 and SDSS J1717$+$5932 was observed
 with the Wide Field Channel of the ACS (ACS/WFC)
 in the F814W band (Proposal ID 9753, PI L. Storrie-Lombardi).
 Since \citet{2010ApJ...711..461S} estimated
 the host-galaxy flux of PG 0844$+$349 in the optical band
 by the image-decomposition method using the same data,
 we adopted their estimate of the $V$-band host-galaxy flux,
 with the additional component of the narrow emission lines
 just being corrected for the difference in the assumed Galactic extinction.

 First we used the two-dimensional image-decomposition program GALFIT
 \citep{2002AJ....124..266P}
 to fit the HST image of the target with the components of
 a PSF for the AGN nucleus,
 the host galaxy, the field galaxies and stars if required,
 and the background-sky contribution.
 We generated the PSF image using the Tiny Tim package
 \citep{1995ASPC...77..349K} version 7.5,
 which models the HST optics and the installed cameras,
 and we modeled the AGN nucleus and the field stars
 with a Gaussian profile
 for which the full width at half maximum was adjusted
 to give the best fit around the AGN nucleus and the field stars.

 Next, we applied aperture photometry using the same size of
 aperture and sky annulus as for the nucleus-free target image
 that is made by subtracting the best-fit PSF component
 from the HST image of the target.
 We analyzed all the unsaturated exposures in the archive
 independently to obtain the host-galaxy flux,
 and we calculated the ensemble standard deviations
 as the errors for PG 0953$+$414 and PG 1613$+$658.
 However, there was only one exposure in the archive
 for SDSS J1717$+$5932, so we calculated the error of
 the host-galaxy flux using IRAF, based on
 background fluctuations and photon noise.

 Finally, we converted the resulting host-galaxy fluxes in the HST bands 
 to those for the optical band used for the monitoring observations.
 For PG 0953$+$414 and PG 1613$+$658,
 we converted the flux in the F550M band to that in the $R$ band
 based on the SWIRE 5 Gyr old elliptical template at the target redshift.
 For SDSS J1717$+$5932,
 we first converted the flux in the F814W band to that in the $I$ band
 and then converted it to that in the $R$ band based on
 linear regression of the multi-epoch flux data in the $R$ and $I$ bands.
 The estimated host-galaxy fluxes and their errors are listed
 in Table \ref{tab:hostflux}.

 \subsubsection{Adopted Host-galaxy Fluxes}
 We examined the three methods for estimating the host-galaxy fluxes
 by comparing the resulting host-galaxy fluxes for the same target
 obtained by two different methods.

 There were four targets for which the host-galaxy fluxes
 were obtained both by the FVG method and by the image-decomposition method.
 The estimates of both methods were consistent to
 within the errors for three targets:
 PG 0844$+$349, PG 0953$+$414, and SDSS J1717$+$5932.
 For PG 1613$+$658, the FVG method estimated
 a slightly negative value for the host-galaxy flux, which is unphysical.

 There were 15 targets for which the host-galaxy fluxes
 were obtained both by the FVG method and by the spectral decomposition method.
 For three of these targets,
 the estimates by the FVG method showed a negative value
 or exceeded the minimum fluxes during the monitoring observations.
 Except for these unphysical estimates by the FVG method,
 we compared those two methods for the remaining 12 targets,
 as presented in Figure \ref{fig:qsfit_fvg}.
 We found that the estimates by both methods were usually
 consistent to within the errors, although the FVG estimates
 generally have larger errors.
 In addition, the FVG estimates seemed to be
 systematically larger than those obtained by the spectral decomposition method.
 They differed by a factor of $1.11\pm 0.04$,
 based on the linear regression analysis.
 We note here that the spectroscopic aperture for the SDSS spectra
 of those targets was $\phi = 3.0$ arcsec in diameter,
 which is slightly smaller than the photometric aperture
 of this study, so we suspect that the host-galaxy fluxes based on
 the SDSS DR10 spectra may be slightly underestimated.

 From these considerations,
 we adopted the host-galaxy fluxes based on the spectral decomposition
 method (17 targets) and the image-decomposition method (five targets)
 when these estimates were available.
 We consider the factor of $1.11$ as a correction factor for
 the possible aperture loss, and we increased the host-galaxy fluxes
 from the spectral decomposition method by that factor\footnote{
 Since the host-galaxy flux for SDSS J1718$+$5313 by the spectral
 decomposition method was very close to the minimum flux
 during the monitoring observation,
 the aperture correction was ignored.
 }.
 We incorporated the uncertainty in the correction factor
 into the errors of the host-galaxy fluxes by root-sum-square.
 We adopted the host-galaxy fluxes obtained from the FVG method
 for the six targets for which they were the only estimates available.
 We found the value of the best estimate with the lower 1$\sigma$ error
 subtracted to be negative for three of these targets,
 and we replaced the lower 1$\sigma$ error
 by the best estimate itself for them.
 We did not estimate the host-galaxy fluxes for the remaining four targets,
 for which we did not obtain a dust-reverberation lag.
 The adopted host-galaxy fluxes and their errors are also listed
 in Table \ref{tab:hostflux}.

 \subsection{Optical Luminosities of the Targets} \label{subsec:agnlum}
 We obtained the representative optical fluxes of the targets
 during the monitoring period as follows.
 First, we linearly interpolated and regularly resampled the light curve.
 Then we averaged the resampled fluxes 
 in order to avoid a possible bias caused by
 the uneven sampling of the observations.
 Finally, we subtracted the host-galaxy flux
 from the average of the resampled fluxes
 to obtain the representative flux of the target.
 In the case of SDSS J1720$+$6128,
 the emission in the rest-frame $V$ band
 is in fact closer to that in the observer-frame $R$ band,
 although the monitoring observations for the dust reverberation
 were conducted in the $I$ band
 (owing to a mistake in the filter selection at the start of observation).
 We therefore converted the $I$-band AGN flux into an $R$-band flux
 based on the slope of the linear regression
 of the multi-epoch flux data in the $R$ and $I$ bands.

 We estimated the $V$-band luminosity of the target from
 the representative optical flux and the luminosity distance
 calculated from the target redshift.
 We used the local flow-corrected redshift of the Virgo infall
 $+$ Great Attractor $+$ Shapley supercluster 
 obtained from the NED for the calculation of luminosity-distance,
 but this correction was very small at large redshifts.
 We corrected the remaining difference in
 the rest-frame wavelength of the observing band
 from the $V$ band by assuming
 a power-law spectrum with $\alpha_{\nu}=-0.3$
 for the AGN continuum emission, as assumed in QSFit.
 The optical fluxes and luminosities of the targets
 are listed in Table \ref{tab:optfL}.

 \subsection{Correlation with the Optical Luminosity}\label{subsec:rLopt}

 In Figure \ref{fig:rLopt},
 we plot the dust-reverberation radius obtained by the MAGNUM project
 against the $V$-band luminosity,
 together with the reverberation radius of the BLR for comparison.
 For the dust-reverberation lags of the Seyfert galaxies
 presented in \citet{2014ApJ...788..159K},
 we adopted the lag times estimated by JAVELIN
 --- instead of using the CCF analysis --- in accordance with this work,
 and calculated the lag times for $\alpha _{VK}=+0.1$
 (where $\alpha _{VK}$ is the assumed power-law index of
 the accretion-disk emission in optical and near-infrared wavelengths)
 by interpolating the data for $\alpha _{VK}=0.0$ and $+1/3$.
 We recalculated their $V$-band luminosities
 by updating the Galactic extinction.
 The recalculated dust-reverberation lags and the luminosities of
 the Seyfert galaxies from \citet{2014ApJ...788..159K}
 are listed in Table \ref{tab:Seyfert}.
 We obtained the data for the BLR from
 the reverberation lags of the Balmer emission lines
 (mostly H$\beta $) compiled by \citet{2013ApJ...767..149B},
 who carefully estimated the optical luminosities
 of their targets using the image-decomposition method
 with the HST images.

 As shown in Figure \ref{fig:rLopt},
 we found the radius of the innermost dust torus inferred from
 the dust reverberation to be strongly correlated with
 the optical luminosity for both Seyfert galaxies and quasars,
 and there appears to be no offset nor break
 between the radius--luminosity relation
 for the Seyfert galaxies and that for the quasars.
 The luminosity range of the sample was about four orders of magnitude,
 which is about 1.5 orders of magnitude
 more luminous in extent than our previous result \citep{2014ApJ...788..159K}.
 As has been reported for the local Seyfert galaxies
 \citep{2006ApJ...639...46S,2014ApJ...788..159K},
 we again found the dust-reverberation radius
 to be located just outside the reverberation radius
 of the BLR for the quasars as well.
 This suggests that the BLR is restricted inside
 the boundary of the dust torus,
 as expected from the unified scheme of AGNs,
 for all luminosity ranges of Seyfert galaxies and quasars.

 In Figure \ref{fig:zllag}, we plot
 the dust-reverberation lag and the $V$-band luminosity
 of the targets against their redshifts.
 The Seyfert galaxies included among the targets for this study
 are distributed at redshifts $z=0.1$--$0.3$,
 and their dust-reverberation lags and luminosities are located
 in the middle of the dust-reverberation radius--luminosity relation
 for the local Seyfert galaxies.
 In addition, the most luminous targets
 are located at redshifts $z=0.23$ and $z=0.60$,
 and their dust-reverberation lags and luminosities are
 very close to each other.
 These results suggest that
 the redshift correction for the dust-reverberation lags
 and the decomposition of the optical AGN fluxes
 were processed reasonably well.
 On the other hand, we found that the targets
 located at redshifts $z\gtrsim 0.35$
 with luminosities $\lambda L_\lambda (V)\sim 10^{44}$ erg s$^{-1}$
 often had large errors in the measured lag times.
 Since they are faint among the targets,
 the photometric errors (especially in the $K$ band) were relatively large.
 Then, the errors in the measured lag times would be large
 unless the amplitude of the flux variations were significantly large.

 We then applied regression analysis
 to the radius--luminosity relation for the innermost dust torus.
 The dust-reverberation lags and the luminosities of the Seyfert
 galaxies and quasars were fitted by the linear equation
 \begin{equation}
  \label{eqn:rL}
 \log r = \alpha + \beta \log \left(\frac{L}{L_0}\right) \, .
 \end{equation}
 We ignored the data for the three targets
 for which the lag times were a lower limit or uncertain
 (the open triangle and circles in Figure \ref{fig:rLopt}),
 and we also ignored the data for the three Seyfert galaxies
 for which the lag times estimated by JAVELIN were different from
 those estimated by the CCF analysis by a factor of $\sim 1.5$ or larger
 (the open boxes in Figure \ref{fig:rLopt}).
 As a result, we used the data for 36 AGNs
 (22 quasars and 14 Seyfert galaxies) for the regression analysis.
 As shown in Figure \ref{fig:rLopt},
 more than half of the ignored targets were in fact consistent with
 the radius--luminosity relation for the 36 selected targets.
 In addition,
 although the dust-reverberation radius of PG 0844$+$349
 inferred from the lag time associated with the peak with the shorter lag
 in the CCF
 was significantly small,
 that for the peak with the longer lag in the CCF was consistent with
 the radius--luminosity relation for the 36 selected targets.
  
 Although the errors in the optical luminosities in Table \ref{tab:optfL}
 were sometimes estimated to have very small values ($\lesssim 0.01$ dex),
 they are unreasonable because possible errors
 caused by the uncertainties in the fitting methods
 --- such as the spectra of the AGN and the host galaxy, etc. ---
 were not incorporated.
 Therefore, we added a moderate fractional error
 (10\%, $\sim 0.04$ dex) to the error
 in the optical luminosity by root-sum-square.
 In addition, we incorporated the peculiar velocity dispersion
 $\sigma _{\rm pec}=300$ km s$^{-1}$ into the luminosity error,
 as has been done by \citet{2014ApJ...788..159K}.
 We then fitted Equation \ref{eqn:rL} to the data,
 adding an error $\sigma_{\rm add}$ to the error of
 the dust-reverberation radius by root-sum-square,
 so that the reduced $\chi ^{2}$ reached unity.
 We adjusted the value of $L_0$ so that
 the errors of $\alpha $ and $\beta $ approximately
 did not correlate with each other.
 For calculation of the regression, we used
 the Marquardt method for the most generalized multivariate
 least-squares process incorporating normally distributed errors
 in all variables without any approximations
 provided by \citet{1980AJ.....85..177J,1981AJ.....86..149J}.
 
 The best-fit regression lines are presented in Figure \ref{fig:rLopt},
 and the resultant parameters are listed in Table \ref{tab:rL}.
 The best-fit linear regression with the slope fixed at $\beta=0.5$,
 which is expected from the radiation equilibrium of dust
 \citep{1987ApJ...320..537B},
 is
 \begin{equation}
  \label{eqn:rL0.5}
 \log r \, [{\rm pc}]\, = (-1.021\pm 0.025) + 0.5 \log \left(\frac{L}{10^{43.7}\, {\rm erg}\, {\rm s}^{-1}}\right) \, .
 \end{equation}
 This is consistent with the result for the Seyfert galaxies
 presented in \citet{2014ApJ...788..159K},
 as no offset nor break is found between the radius--luminosity relation
 for the Seyfert galaxies and that for the quasars.

 When the slope $\beta$ is set as a free parameter,
 the best-fit linear regression is
 \begin{equation}
  \label{eqn:rLfree}
 \log r \, [{\rm pc}]\, = (-1.021\pm 0.023) + (0.424\pm 0.026) \log \left(\frac{L}{10^{43.7}\, {\rm erg}\, {\rm s}^{-1}}\right) \, .
 \end{equation}

 In addition to the primary method for the regression analysis,
 we calculated the regression using two different methods
 to examine possible uncertainties introduced by the fitting methods.
 One is FITEXY estimator \citep{1992nrfa.book.....P}
 for a bivariate least-squares process incorporating errors
 in dependent and independent variables.
 We used the MPFITEXY routine \citep{2010MNRAS.409.1330W},
 which depends on the MPFIT package \citep{2009ASPC..411..251M} in IDL.
 The other is a Bayesian linear regression routine linmix\_err in IDL
 developed by \citet{2007ApJ...665.1489K}, which incorporates
 normally distributed errors in dependent and independent variables.
 It models the distribution of the independent variable
 by a mixture of Gaussian functions,
 and we set the parameter {\tt NGAUSS}$=1$ for the number of them.
 Since it fits the data using a Markov Chain Monte Carlo method,
 we determined the best estimate and the $\pm 1\sigma $ errors
 for the linear-regression parameters
 from the median, 15.9th percentile, and 84.1th percentile
 of their likelihood distributions.
 Both methods simultaneously fit the intrinsic scatter,
 which corresponds to $\sigma_{\rm add}$ for the primary method.
 We calculated the regression only for the slope $\beta$ being set
 as a free parameter by these methods,
 and the results are also listed in Table \ref{tab:rL}.
 The results of the regression analysis by different methods
 agree well with each other, thus they are thought to be robust.
 
 We find that the best-fit slope is smaller than $0.5$
 at about the $3\sigma$ level.
 A number of possible mechanisms may make the slope smaller.
 For example, \citet{2010ApJ...724L.183K,2011ApJ...737..105K}
 showed that
 anisotropic illumination of the accretion disk
 makes the dust-sublimation radius smaller in the equatorial plane
 than in the polar directions.
 If the fraction of the geometrical thickness relative to the radius
 is smaller for more luminous AGNs as suggested by many studies
 \citep[e.g.][]{1991MNRAS.252..586L,2003ApJ...598..886U,
  2005MNRAS.360..565S,2007A&A...468..979M,2014MNRAS.437.3550M},
 their model predicts that 
 the inner radius of the dust torus
 relative to the square root of the luminosity
 is smaller for more luminous AGNs.
 
 On the other hand,
 the inner dust torus was found not to contract
 immediately after the propagation 
 of a drop in luminosity of the accretion disk
 \citep{2009ApJ...700L.109K, 2010ApJ...715..736P,
 2013ApJ...775L..36K,2014AstL...40..527O,2019arXiv190408946K}.
 If the response timescale to complete the contraction
 is less than a few years, as reported by
 \citet{2009ApJ...700L.109K} and \citet{2019arXiv190408946K},
 an AGN whose accretion disk has already dimmed
 but for which the dust torus still remains large
 would be observed more frequently in less luminous AGNs,
 in which the timescale for flux variation is expected to be small.
 In these objects, the inner radius of the dust torus on average
 would be larger for less luminous AGNs.
 \footnote{\citet{2009ApJ...700L.109K} also reported that
 the inner dust torus did not expand immediately after
 the propagation of an increase in luminosity of the accretion disk.
 If the clumpy dust clouds located within the sublimation radius
 can survive temporarily,
 an AGN whose accretion disk has already brightened
 but for which the dust torus still remains small
 may be observed more frequently in less luminous AGNs.
 }
 
 A possible systematic change in the spectral shape
 of the UV--optical continuum emission of the accretion disk
 with the luminosity
 would also change the dust-sublimation radius from the correlation
 with the square root of the $V$-band luminosity,
 because the integrated input energy for dust grains
 normalized by the $V$-band flux depends on the spectral shape
 of the illumination
 \citep[][]{1987ApJ...320..537B,2014ApJ...784L..11Y}.
 The luminosity dependence of the spectral energy distributions (SEDs) of AGNs
 has been studied by many authors,
 but \citet{2013ApJS..206....4K} reported
 a somewhat complicated behavior in the UV--optical continuum,
 with more luminous AGNs tending to show
 a bluer optical continuum and a redder far-UV continuum.
 The influence of the luminosity-dependent
 UV--optical continuum on the radius--luminosity relation
 is thus not simple.
 
 However, we suspect that it is still uncertain
 whether the slope $\beta $ is significantly smaller than 0.5.
 First, the number of the targets at both ends of the luminosity range
 is still limited.
 In addition, there are possibly observational biases
 against the largest and smallest lags 
 due to the limited monitoring span and cadence,
 which --- in combination with the scatter in the correlation ---
 would make the slope smaller than its intrinsic value.
 Moreover, systematic uncertainties in
 the subtraction of the accretion-disk component in the near-infrared
 and the wavelength dependence of the dust-reverberation lag
 may also influence the slope measurement,
 because more luminous targets are located at larger redshifts,
 as shown in Figure \ref{fig:zllag}.
 In order to clarify this point,
 it is desirable to apply the dust reverberation to more AGNs
 that cover wider ranges of both redshift and luminosity
 by longer-term monitoring observations of luminous quasars
 and higher-cadence ones for less luminous Seyfert galaxies.

 3C 273 ($z=0.158$) is the most luminous quasar for which
 a dust-reverberation lag in the $K$ band has been obtained
 \citep{2008A&A...486..411S}.
 Its lag time was measured to be
 $\Delta t \sim 1$ yr ($c\Delta t \sim 0.3$ pc),
 and its optical luminosity was approximately 
 $\lambda L_{\lambda }(V)\sim 10^{46}$ erg s$^{-1}$.
 Therefore, its dust-reverberation radius is smaller
 than the extrapolation of the radius--luminosity relation
 presented in Figure \ref{fig:rLopt}.
 This suggests either that 
 the increase in the radius of the innermost dust torus
 saturates in the most luminous quasars
 or that the slope $\beta $ is significantly smaller than $0.5$.
 However, \citet{2008A&A...486..411S} indicated that
 the optical continuum flux is dominated by
 an ``$R$ component'' \citep{1998A&A...340...47P}
 that may be synchrotron emission,
 and the luminosity of the accretion-disk component may thus be
 an order of magnitude smaller than the total $V$-band luminosity,
 as shown in their Figure 5.
 If the $R$ component is highly anisotropic and does not illuminate
 the dust torus, only the accretion-disk emission would contribute
 input energy to the radiation equilibrium of the dust.
 The dust-reverberation radius of this quasar is consistent with
 the radius--luminosity relation presented here
 when we use the $V$-band luminosity of the accretion-disk component.
 The same would hold for the interferometric dust radius
 of 3C 273
 \citep{2011A&A...527A.121K,2018Natur.563..657G,2019arXiv191000593G},
 which may influence the discussion
 of the radius--luminosity relation for the interferometric dust radius
 and the difference from the dust-reverberation radius
 \citep{2011A&A...527A.121K,2014ApJ...788..159K,2019arXiv191000593G}.

 \subsection{Correlation with Isotropic Luminosity Indicators}\label{subsec:rLother}

 We examined the correlation between the dust-reverberation radius
 and the isotropic luminosity indicators of AGNs.
 Although these luminosity indicators are not dominant radiation
 sources for heating dust in the torus,
 they are the luminosity indicators for AGNs
 that are far less obscured by dust.
 Correlations with these indicators may thus be useful
 for studying the inner structures of obscured AGNs,
 such as type 2 AGNs and ultraluminous infrared galaxies.
 
 We present here the correlations with
 three isotropic luminosity indicators for AGNs:
 the hard X-ray luminosity,
 the luminosity of [O{\small IV}] 25.89 $\mu$m emission line
 in the mid-infrared,
 and the luminosity of the mid-infrared continuum emission.
 The former two are updates of the results in \citet{2014ApJ...788..159K},
 while the last one is a newly examined correlation.
 The correlation with X-ray luminosity has also been examined
 by \citet{2011A&A...531A..99T} and \citet{2016ApJ...821...15F}
 based on our early data on dust reverberation
 \citep{2006ApJ...639...46S}.
 We took the hard X-ray luminosities of the target AGNs from
 the 14--195 keV luminosities of AGNs by \citet{2017ApJS..233...17R}
 based on the 70-month Swift/BAT catalog \citep{2013ApJS..207...19B}.
 We obtained the [O{\small IV}] emission-line luminosities
 from a number of previous studies
 \citep{2007ApJ...671..124D,2008ApJ...682...94M,2009ApJ...698..623D,
 2010ApJ...723..409G,2010ApJ...725.2381L,
 2010ApJ...709.1257T,2010ApJ...716.1151W,2018MNRAS.478.4068L}.
 We employed the mid-infrared continuum luminosities
 of the nuclear components at the wavelength $\lambda = 12$ $\mu$m
 measured by \citet{2014MNRAS.439.1648A}
 based on the mid-infrared images at subarcsecond resolution
 obtained by ground-based 8 m class telescopes.
 The isotropic luminosity indicators of the target AGNs
 are listed in Table \ref{tab:isoL}.

 Figures \ref{fig:rLhx}--\ref{fig:rLmir} present
 the dust-reverberation radius
 plotted against the isotropic luminosity indicators,
 together with the reverberation radius of the BLR
 obtained by \citet{2013ApJ...767..149B} for comparison.
 We applied the regression analysis to those data
 in the same way as for the $V$-band luminosity.
 The best-fit regression lines are presented
 in Figure \ref{fig:rLhx}--\ref{fig:rLmir},
 and the resulting parameters are listed in Table \ref{tab:rL}.
 Since only a few quasars have data for
 those isotropic luminosity indicators,
 the increases in the quantity of the data and in the luminosity range
 were limited. 
 The best-fit regression for
 the hard X-ray luminosity and the [O{\small IV}] emission-line luminosity
 are consistent with those of \citet{2014ApJ...788..159K}.
 Although the quantity of data is small,
 we found that the dust-reverberation radius was
 tightly correlated with the mid-infrared continuum luminosity
 for the Seyfert galaxies, as presented
 in Figure \ref{fig:rLmir} and Table \ref{tab:rL}.
 The slopes of the best-fit linear regressions
 for all isotropic luminosity indicators
 are consistent with $\beta = 0.5$
 within 1$\sigma $ error when it was set as a free parameter.
 This is partly because of the large errors in slope
 caused by the limited quantity of data and limited luminosity range.
 Further observations of the isotropic luminosity indicators
 for more luminous AGNs will enable us to examine more details
 of the correlations between the dust-reverberation radius
 and the isotropic luminosity indicators.
 We note that the dust-reverberation radius of PG 0844$+$349
 inferred from the lag time associated with the peak with the shorter lag
 in the CCF
 is consistent with the radius--luminosity relation for
 the [O{\small IV}] emission-line and the mid-infrared continuum luminosities,
 in contrast to the optical $V$-band luminosity.
 Even from the radius--luminosity relations,
 it is still uncertain which lag time is favored to represent
 the radius of the innermost dust torus for PG 0844$+$349.

 \section{Constraining Cosmic Expansion Using Dust-reverberation Distance} \label{subsec:hubble}

 \citet{2014ApJ...784L..11Y} formulated an equation
 to estimate the luminosity distance of an AGN
 from the dust-reverberation lag and the optical flux,
 based on a model of the radiation equilibrium
 of dust grains in the innermost dust torus,
 and they applied the formula to the data for Seyfert galaxies
 \citep{2014ApJ...788..159K} to obtain the Hubble constant.
 We apply it here to the data for the Seyfert galaxies and quasars
 at redshifts $z<0.6$ to plot a Hubble diagram and
 examine the cosmological models.

 Since we selected the monitoring optical bands
 from among the $V$, $R$, and $I$ bands
 so as to be close to the rest-frame $V$ band,
 we slightly modified Equation 2 of \citet{2014ApJ...784L..11Y}
 to give the dust-reverberation distance $d$ in megaparsecs as follows:
 \begin{equation}
  \label{eqn:drd}
 d=\Delta t \times 10^{0.2(m'_{X}-k'-25+g)}\, ,
 \end{equation}
 where
 \begin{equation}
 m'_{X}= -2.5 \log \left(\frac{f_{\nu }(\nu _{X})}{f_{\nu ,0}(\nu _{V})}\right)\, ,
 \end{equation}
 and
 \begin{equation}
 k'=-2.5 \log (1+z) -2.5 \alpha _{\nu } \log \left(\frac{(1+z)\nu _{X}}{\nu _{V}}\right)\, .
 \end{equation}
 Here, $X$ represents the monitoring optical bands ($V$, $R$, or $I$),
 $\nu _{X}$ is the effective frequency of the $X$ band,
 $f_{\nu }(\nu _{X})$ is the $X$-band flux
 corrected for Galactic extinction,
 and $f_{\nu ,0}(\nu _{V})$ is the flux for $m_{V}=0$.
 We assumed $\alpha _{\nu }=-0.3$ for the power-law index
 for the optical continuum emission of the AGNs
 around the effective frequency of the $V$ band (rest frame).
 We used $g=10.60$ for the distance-calibration factor
 \citep{2014ApJ...784L..11Y}.
 
 Figure \ref{fig:hubble} plots the dust-reverberation distance
 against the redshift for the 36 Seyfert galaxies and quasars
 in our sample.
 As shown in Figure \ref{fig:hubble}, the cosmological model
 with current standard estimates 
 ($H_0=73$ km s$^{-1}$ Mpc$^{-1}$, $\Omega_0=0.27$, $\lambda_0=0.73$)
 is consistent with the data.
 The best-fit cosmological parameters assuming a flat universe are
 $H_0=68.4$ km s$^{-1}$ Mpc$^{-1}$ and $\Omega_0 =0.44$ ($\lambda_0=0.56$),
 with a $\pm 1\sigma $ confidence interval of $\Omega_0 =0.14$--$0.82$
 when $H_0$ is fixed at the best-fit value.
 As a result, an accelerated universe is preferred
 from the dust-reverberation distances of the AGNs.
 We note that
 the deceleration parameter of the universe is sensitive
 to the difference between
 the luminosity distances of local AGNs
 and those of more distant ones.
 Therefore,
 precise observations of local AGNs are
 as important as those of AGNs at large redshifts.
 In contrast, since the uncertainty in $g$ transferred from
 the uncertainties in the radiation-equilibrium model of the dust
 only scales the dust-reverberation distance
 by the same amount at all redshifts,
 it does not contribute to errors in measuring
 the acceleration of the cosmic expansion.
 
 Unfortunately, the cosmological parameters ($\Omega_0, \lambda_0$)
 are not tightly constrained
 by the dust-reverberation distance in this study,
 partly because the number of targets with precisely measured
 dust-reverberation lags located at large redshifts ($z\gtrsim 0.35$)
 is still small.
 More monitoring observations
 with high cadence, long term, and good photometric accuracy
 will improve the constraints on the cosmological parameters.

 Moreover, understanding the origins of scatter in
 the correlation between the dust-reverberation radius
 and the luminosity are important for developing appropriate ways
 to improve the constraints on the cosmological parameters.
 We note that the correlation with the mid-infrared continuum luminosity
 is rather better than that with the optical luminosity,
 which suggests that further improvement of the radius--luminosity correlation,
 and thus of the dust-reverberation distance, is promising.
 The delayed response of the contraction (or expansion)
 of the inner dust torus after the propagation of a change in luminosity,
 and the target-to-target variation of
 the SED of the accretion-disk emission,
 are possible origins of scatter in the correlation.
 Simultaneously obtained UV--optical--near-infrared continuum spectra
 are expected to be useful to compensate for these problems.
 When the dust is located away from the global correlation,
 the dust temperature determined by radiation equilibrium
 would be different from the typical value,
 which can be observed in the SED of the near-infrared dust emission.
 Then, two parameters in the radiation-equilibrium model of dust
 for the distance-calibration factor $g$
 ---
 the SED of the UV--optical continuum emission of the accretion disk
 and the dust temperature
 ---
 can be constrained observationally target-by-target
 to improve the dust-reverberation distance
 \citep{2014ApJ...784L..11Y}.
 This spectral information is also useful for subtracting 
 the accretion-disk component from the near-infrared fluxes,
 which is important for extending the sample of dust-reverberation AGNs
 to much larger redshifts.
 As \citet{2014ApJ...788..159K} examined a possible systematic difference
 from the global radius--luminosity relation for the innermost dust torus
 depending on the Eddington ratio and the viewing angle,
 understanding the geometry and structure in the central region
 of the dust torus and their possible influence on the dust-reverberation lag
 are also important for obtaining an accurate dust-reverberation distance.
 
 At the end of this section,
 we discuss possible sources of systematic uncertainties
 in constraining the cosmic expansion.
 One is the uncertainty in the correction for
 the wavelength dependence of the dust-reverberation lag,
 because a different redshift correction to the lag time produces
 a systematic difference in the dust-reverberation distance
 that depends on redshift,
 which will alter the constraints on the cosmological parameters.
 Another is the uncertainty in the slope of the radius--luminosity relation.
 If it is different from 0.5,
 Equation \ref{eqn:drd} estimates systematically different distances
 that depend on the luminosity,
 and this will also alter the constraints on the cosmological parameters
 when the luminosity distribution of the targets changes with redshift.
 Understanding
 more details about the wavelength dependence of the dust-reverberation lag
 and the correlation of the dust-reverberation radius with the luminosity
 is important for constraining the cosmological parameters accurately.
 On the other hand,
 monitoring observations of less luminous targets at large redshifts
 and at longer wavelengths ($3$--$5$ $\mu$m)
 and those of luminous targets in the local universe
 at shorter wavelengths ($H$ and $J$ bands)
 will reduce these systematic uncertainties observationally.
 \citet{2019arXiv190911101L}
 actually demonstrated the feasibility of dust reverberation
 at longer wavelengths ($3$--$5$ $\mu$m) for quasars
 at large redshifts ($z\lesssim 0.5$).

 \section{Summary} \label{sec:summary}
 We have presented the results of a dust-reverberation survey
 of quasars at redshifts $z<0.6$ obtained by the MAGNUM project.
 We found a delayed response of the $K$-band flux variation
 after the optical flux variation for 25 out of 31 targets.
 For more than half of the remaining six targets,
 a low signal-to-noise ratio and probable systematic errors
 in the $K$-band photometry
 make a possible delayed response uncertain.
 Therefore, we consider dust reverberation to be ubiquitous,
 not only for Seyfert galaxies but also for quasars.

 We measured the dust-reverberation lag using JAVELIN,
 fitting the data for the $K$-band flux variations
 with the delayed response of the dust-torus emission
 superimposed on the accretion-disk component
 that is proportional to the optical flux variation.
 We succeeded in estimating the dust-reverberation lag
 for 22 out of the 25 targets with a delayed response,
 which include the most distant AGN with a dust-reverberation lag
 and the AGN with the largest dust-reverberation lag in the $K$ band
 measured so far.
 Combined with a selected sample of 14 out of
 the 17 nearby Seyfert galaxies studied by \citet{2014ApJ...788..159K},
 we provide the largest homogeneous collection
 of $K$-band dust-reverberation data for 36 type 1 AGNs.

 Since the flux obtained by aperture photometry contains
 a significant contribution from the host-galaxy starlight,
 we estimated the optical luminosity of the AGN component
 by three different methods: spectral decomposition,
 the FVG method, and image decomposition.
 We then examined the correlation between the reverberation radius
 for the innermost dust torus and the optical luminosity.
 We found that they are strongly correlated,
 as is already known for the Seyfert galaxies,
 and the best-fit radius--luminosity relation is
 $\log r\, {\rm [pc]}=-1.021 + 0.5\log (\lambda L_{\lambda}(V)/10^{43.7}\, {\rm erg\, s^{-1}})$
 or
 $\log r\, {\rm [pc]}=-1.021 + 0.424\log (\lambda L_{\lambda}(V)/10^{43.7}\, {\rm erg\, s^{-1}})$
 when the slope of the relation is fixed at $0.5$ or freed, respectively.
 The latter slope is about $3\sigma $ smaller than
 the value of 0.5 that is expected from the radiation equilibrium of dust.
 However, it is still uncertain whether the slope
 is significantly smaller than $0.5$.
 We found that the dust-reverberation radius to be located
 just outside of the reverberation radius of the BLR,
 even for quasars, as expected from the unified scheme of AGNs.
 We also presented the radius--luminosity relations
 for the isotropic luminosity indicators,
 which will be applicable for obscured AGNs.
 We updated the relations for the hard X-ray (14--195 keV) and
 the [O{\small IV}] $\lambda $25.89 $\mu$m luminosities
 from a previous study \citep{2014ApJ...788..159K},
 and we presented the correlation with
 the mid-infrared (12 $\mu$m) continuum luminosity for the first time.

 Using this homogeneous data collection,
 we estimated the luminosity distances of the 36 AGNs
 based on their dust-reverberation lags \citep{2014ApJ...784L..11Y},
 and we found that the data for the redshift--luminosity distance diagram 
 are consistent with current standard estimates for the cosmological parameters.
 We note that the uncertainties in the calibration factor
 for the dust-reverberation distance do not produce errors
 in measuring the acceleration of the cosmic expansion.
 Unfortunately, the cosmological parameters are not tightly constrained
 by the dust-reverberation distance in this study.
 In addition to the random errors,
 we discussed possible sources of systematic uncertainties
 in the study of cosmology based on the dust-reverberation distances.

Ongoing and future dust-reverberation mapping surveys
for a large number of distant AGNs with different luminosities
by large telescopes
(VISTA, H\"{o}nig et al. 2017; LSST, H\"{o}nig 2014; TAO, Yoshii et al. 2016),
as well as multi-wavelength dust-reverberation mapping
with high cadence and good photometric accuracy for local AGNs
\citep[e.g.,][]{2019MNRAS.489.1572L},
will enable us to investigate more details of the geometry and structure 
in the central region of the dust torus,
and to improve constraints on the cosmic expansion
based on the dust-reverberation distances.
Results from dust reverberation for type 1 AGNs,
combined with less obscured emission such as X-rays,
are applicable to the study of obscured AGNs
\citep[e.g.,][]{2015ApJ...802...98M}.
Precise high-resolution X-ray spectroscopy that is planned for
the next Japanese X-ray satellite XRISM \citep{2018SPIE10699E..22T}
will enable us to study the dynamics in the central region of AGNs,
as demonstrated by the results on NGC 1275 obtained by the Hitomi satellite
\citep{2018PASJ...70...13H}.

  
\acknowledgments

We thank the staff at the Haleakala Observatories for their help
with facility maintenance, and Mitsuru Kokubo for useful discussion.
This research has been supported partly by
the Grants-in-Aid of Scientific Research (10041110, 10304014, 11740120,
12640233, 14047206, 14253001, 14540223, 16740106, and 16H02162)
and the COE Research (07CE2002) of the Ministry of Education,
Science, Culture, and Sports of Japan.

\software{
JAVELIN software \citep{2011ApJ...735...80Z},
PYCCF software \citep{2018ApJ...857...86S},
QSFit software \citep{2017MNRAS.472.4051C},
Tiny Tim package \citep[v7.5][]{1995ASPC...77..349K},
MPFITEXY \citep{2010MNRAS.409.1330W},
MPFIT package \citep{2009ASPC..411..251M},
linmix\_err \citep{2007ApJ...665.1489K}.
}


\clearpage
\begin{deluxetable*}{lllrcccrrl}
\tablecaption{List of Targets and Monitoring Parameters\label{tab:objdata}}
\tabletypesize{\scriptsize}
\tablewidth{0pt}
\tablehead{
\colhead{Name} & \colhead{R.A.} & \colhead{Decl.} & \colhead{Redshift\tablenotemark{\scriptsize a}} &
 \multicolumn2c{Band\tablenotemark{\scriptsize b}} & \colhead{Observing Period} &
 \colhead{$n_{\rm obs}$\tablenotemark{\scriptsize c}} &
 \colhead{$t_{\rm int}$\tablenotemark{\scriptsize d}} & \colhead{Full Name\tablenotemark{\scriptsize e}}
}
\startdata
SDSS J0007$-$0054  & 00 07 29.990 & $-$00 54 28.01 & 0.1452 & $R$ & $K$ & 2001.11.20--2007.08.18 & 131 &   6 & SDSS J000729.99$-$005428.0 \\
LBQS 0023$+$0228   & 00 26 21.94  & $+$02 44 41.8  & 0.2360 & $R$ & $K$ & 2003.06.29--2007.08.02 &  56 &  12 & \\
SDSS J0207$-$0048  & 02 07 24.177 & $-$00 48 41.41 & 0.3728 & $I$ & $K$ & 2001.11.24--2007.08.06 &  73 &  12 & SDSS J020724.17$-$004841.4 \\
SDSS J0315$+$0012  & 03 15 42.642 & $+$00 12 28.75 & 0.2073 & $R$ & $K$ & 2001.11.13--2007.08.17 & 112 &   7 & SDSS J031542.64$+$001228.7 \\
PG 0844$+$349      & 08 47 42.465 & $+$34 45 04.40 & 0.0640 & $V$ & $K$ & 2003.02.23--2007.05.20 &  25 &  34 & \\
SDSS J0943$-$0043  & 09 43 35.613 & $-$00 43 22.05 & 0.2714 & $R$ & $K$ & 2002.01.15--2007.05.30 &  67 &  14 & SDSS J094335.61$-$004322.0 \\
PG 0953$+$414      & 09 56 52.392 & $+$41 15 22.25 & 0.2341 & $R$ & $K$ & 2003.02.17--2007.06.19 &  12 & 112 & \\
SDSS J0957$-$0023  & 09 57 58.449 & $-$00 23 54.03 & 0.5956 & $I$ & $K$ & 2002.01.11--2007.04.03 &  16 &  87 & SDSS J095758.44$-$002354.0 \\
2QZ J1013$+$0028   & 10 13 55.210 & $+$00 28 50.28 & 0.3305 & $R$ & $K$ & 2002.02.04--2007.05.31 &  52 &  18 & 2QZ J101355.1$+$002849 \\
LBQS 1022$-$0005   & 10 24 50.521 & $-$00 21 02.41 & 0.3203 & $R$ & $K$ & 2003.10.29--2007.06.04 &  22 &  35 & \\
LBQS 1026$-$0032   & 10 29 20.702 & $-$00 47 47.60 & 0.2592 & $R$ & $K$ & 2002.02.15--2007.06.01 &  33 &  33 & \\
2QZ J1032$-$0233   & 10 32 24.642 & $-$02 33 21.66 & 0.5670 & $I$ & $K$ & 2003.11.23--2007.05.22 &  12 &  88 & 2QZ J103224.6$-$023322 \\
SDSS J1044$+$0003  & 10 44 49.281 & $+$00 03 01.22 & 0.4428 & $I$ & $K$ & 2002.01.09--2007.06.24 &  40 &  28 & SDSS J104449.28$+$000301.2 \\
2QZ J1138$-$0131   & 11 38 07.196 & $-$01 31 58.66 & 0.4780 & $I$ & $K$ & 2002.01.15--2007.06.19 &  47 &  24 & 2QZ J113807.1$-$013159 \\
2QZ J1220$-$0119   & 12 20 32.471 & $-$01 19 49.16 & 0.5130 & $I$ & $K$ & 2003.02.09--2007.06.01 &  26 &  31 & 2QZ J122032.4$-$011950 \\
2QZ J1225$-$0101   & 12 25 49.442 & $-$01 01 54.11 & 0.5750 & $I$ & $K$ & 2003.12.22--2007.07.15 &  21 &  38 & 2QZ J122549.4$-$010155 \\
2QZ J1247$+$0025   & 12 47 41.703 & $+$00 25 16.77 & 0.4110 & $I$ & $K$ & 2003.12.24--2007.07.13 &  32 &  25 & 2QZ J124741.6$+$002515 \\
SDSS J1309$-$0015  & 13 09 16.672 & $-$00 15 50.15 & 0.4228 & $I$ & $K$ & 2004.02.02--2007.07.26 &  25 &  26 & SDSS J130916.67$-$001550.1 \\
LBQS 1339$+$0210   & 13 42 16.177 & $+$01 55 18.67 & 0.2728 & $R$ & $K$ & 2003.02.16--2007.08.08 &  32 &  32 & \\
2QZ J1345$-$0231   & 13 45 12.443 & $-$02 31 46.18 & 0.5283 & $I$ & $K$ & 2003.02.11--2007.08.02 &  23 &  46 & 2QZ J134512.4$-$023147 \\
2QZ J1438$-$0116   & 14 38 22.920 & $-$01 16 36.05 & 0.3880 & $I$ & $K$ & 2003.02.11--2007.08.10 &  64 &  14 & 2QZ J143822.8$-$011636 \\
PG 1613$+$658      & 16 13 57.179 & $+$65 43 09.58 & 0.1290 & $R$ & $K$ & 2003.09.04--2007.08.08 &  27 &  39 & \\
SDSS J1717$+$5932  & 17 17 47.566 & $+$59 32 58.08 & 0.2483 & $R$ & $K$ & 2003.02.28--2007.08.18 &  75 &  12 & SDSS J171747.55$+$593258.0 \\
SDSS J1718$+$5313  & 17 18 32.858 & $+$53 13 04.67 & 0.1917 & $R$ & $K$ & 2003.02.26--2007.08.18 &  92 &   9 & SDSS J171832.85$+$531304.7 \\
SDSS J1720$+$6128  & 17 20 59.476 & $+$61 28 11.79 & 0.2365 & $I$ & $K$ & 2003.02.08--2007.08.09 &  48 &  23 & SDSS J172059.45$+$612811.7 \\
SDSS J1723$+$5400  & 17 23 00.547 & $+$54 00 55.84 & 0.4806 & $I$ & $K$ & 2002.08.23--2007.07.19 &  32 &  36 & SDSS J172300.53$+$540055.7 \\
SDSS J1724$+$6036  & 17 24 46.401 & $+$60 36 19.63 & 0.3715 & $I$ & $K$ & 2003.02.17--2007.07.26 &  32 &  35 & SDSS J172446.40$+$603619.6 \\
IRAS F21256$+$0219 & 21 28 12.32  & $+$02 32 31.9  & 0.2570 & $R$ & $K$ & 2001.10.11--2007.08.02 &  70 &  17 & \\
RX J2138$+$0112    & 21 38 18.971 & $+$01 12 22.45 & 0.3441 & $I$ & $K$ & 2001.08.03--2007.08.18 &  67 &  16 & RX J2138.2$+$0112 \\
RX J2156$+$1426    & 21 56 42.61  & $+$14 26 34.4  & 0.3560 & $I$ & $K$ & 2001.07.27--2007.08.18 & 134 &   8 & RX J2156.7$+$1426 \\
SDSS J2326$-$0030  & 23 26 40.012 & $-$00 30 41.41 & 0.5820 & $I$ & $K$ & 2003.06.01--2007.07.22 &  22 &  41 & SDSS J232640.01$-$003041.4 \\
\enddata
\tablecomments{}
\tablenotetext{a}{The heliocentric redshift from the NED.}
\tablenotetext{b}{Observing photometric bands to measure the dust-reverberation lag, column 5 for optical and column 6 for near-infrared.}
\tablenotetext{c}{Number of observations.}
\tablenotetext{d}{Median monitoring interval in unit of days.}
\tablenotetext{e}{Full name for the Sloan Digital Sky Survey (SDSS),
 the 2dF QSO Redshift Survey (2QZ), and ROSAT X-Ray source (RX).}
\end{deluxetable*}

\clearpage
\begin{deluxetable*}{lrrrr}
\tablecaption{Light-curve Data\label{tab:lcdata}}
\tabletypesize{\footnotesize}
\tablewidth{0pt}
\tablehead{
\colhead{Object} & \colhead{Band} & \colhead{Observation Date} & \colhead{Flux} & \colhead{Flux Error} \\
 & & \colhead{(MJD)} & \colhead{($\mu$Jy)} & \colhead{($\mu$Jy)}
}
\startdata
 SDSS J0007$-$0054  & $R$ & 52226.38 & 172.1 & 1.4 \\
                    &     & 52233.35 & 172.6 & 0.8 \\
                    &     & \nodata  & \nodata & \nodata \\
 SDSS J0007$-$0054  & $K$ & 52233.35 & 703.9 & 5.2 \\
                    &     & 52237.34 & 703.9 & 4.5 \\
                    &     & \nodata  & \nodata & \nodata \\
 LBQS 0023$+$0228   & $R$ & 52819.57 & 121.6 & 0.5 \\
                    &     & 52849.56 & 121.7 & 0.8 \\
                    &     & \nodata  & \nodata & \nodata \\
 \nodata            & \nodata & \nodata & \nodata & \nodata \\
\enddata
\tablecomments{Galactic extinction has been corrected for, and the fluxes from the host galaxy have not been subtracted. (This table is available in its entirety in a machine-readable form.)}
\end{deluxetable*}

\clearpage
\begin{deluxetable*}{lccccc}
\tablecaption{Dust-reverberation Lags\label{tab:lagdata}}
\tabletypesize{\footnotesize}
\tablewidth{0pt}
\tablehead{
\colhead{Object} & \colhead{Band\tablenotemark{\footnotesize a}} & \multicolumn{3}{c}{Lag Time\tablenotemark{\footnotesize b}} & \colhead{Lag Time\tablenotemark{\footnotesize c}} \\
 & & \multicolumn{3}{c}{(days)} & \colhead{(days)} \\
 & & \colhead{$\alpha_{\rm OIR}=+1/3$\tablenotemark{\footnotesize d}} & \colhead{$+0.1$\tablenotemark{\footnotesize d}} & \colhead{$0.0$\tablenotemark{\footnotesize d}} & \colhead{$+0.1$\tablenotemark{\footnotesize d}}
 }
 \startdata
 SDSS J0007$-$0054 & $R$, $K$ & $35.9^{+2.8}_{-2.8}$ & $35.9^{+2.8}_{-2.8}$ & $35.9^{+2.7}_{-2.7}$ 	& $36.8^{+2.9}_{-2.9}$ \\
  LBQS 0023$+$0228 & $R$, $K$ & \nodata & \nodata & \nodata 	& \nodata \\
 SDSS J0207$-$0048 & $I$, $K$ & $231.6^{+62.9}_{-62.9}$ & $235.6^{+62.1}_{-62.1}$ & $237.5^{+61.6}_{-61.6}$ 	& $249.5^{+65.8}_{-65.8}$ \\
 SDSS J0315$+$0012 & $R$, $K$ & $76.26^{+0.48}_{-0.57}$ & $76.21^{+0.49}_{-0.60}$ & $76.19^{+0.48}_{-0.57}$ 	& $78.8^{+0.5}_{-0.6}$ \\
     PG 0844$+$349\tablenotemark{\footnotesize e} & $V$, $K$ & $99.7^{+10.9}_{-10.3}$ & $98.1^{+12.6}_{-10.0}$ & $99.3^{+14.9}_{-10.1}$ 	& $99.2^{+12.7}_{-10.1}$ \\
 SDSS J0943$-$0043 & $R$, $K$ & $158.1^{+16.4}_{-14.5}$ & $161.1^{+16.5}_{-15.3}$ & $161.3^{+16.9}_{-15.3}$ 	& $168.2^{+17.2}_{-16.0}$ \\
     PG 0953$+$414 & $R$, $K$ & $485.4^{+35.8}_{-22.3}$ & $545.2^{+48.5}_{-37.0}$ & $575.6^{+47.8}_{-40.4}$ 	& $566.3^{+50.3}_{-38.4}$ \\
 SDSS J0957$-$0023 & $I$, $K$ & $596.6^{+79.5}_{-79.5}$ & $591.6^{+83.0}_{-83.0}$ & $590.4^{+81.9}_{-81.9}$ 	& $643.5^{+90.3}_{-90.3}$ \\
  2QZ J1013$+$0028\tablenotemark{\footnotesize f} & $R$, $K$ & $166.5^{+27.6}_{-27.6}$ & $170.8^{+27.5}_{-27.5}$ & $175.2^{+25.9}_{-25.9}$ 	& $179.8^{+28.9}_{-28.9}$ \\
  LBQS 1022$-$0005 & $R$, $K$ & $217.2^{+13.9}_{-10.9}$ & $221.3^{+13.5}_{-12.6}$ & $223.7^{+12.5}_{-12.5}$ 	& $232.6^{+14.2}_{-13.3}$ \\
  LBQS 1026$-$0032 & $R$, $K$ & $213.1^{+42.2}_{-22.9}$ & $233.9^{+47.7}_{-31.0}$ & $249.7^{+44.2}_{-38.2}$ 	& $243.8^{+49.7}_{-32.3}$ \\
  2QZ J1032$-$0233 & $I$, $K$ & \nodata & \nodata & \nodata 	& \nodata \\
 SDSS J1044$+$0003 & $I$, $K$ & $108.1^{+24.0}_{-20.4}$ & $116.9^{+25.1}_{-22.8}$ & $122.1^{+24.8}_{-22.6}$ 	& $124.9^{+26.8}_{-24.3}$ \\
  2QZ J1138$-$0131 & $I$, $K$ & $85.3^{+26.4}_{-30.9}$ & $90.4^{+22.0}_{-23.3}$ & $92.5^{+20.9}_{-22.4}$ 	& $97.0^{+23.6}_{-25.0}$ \\
  2QZ J1220$-$0119 & $I$, $K$ & \nodata & \nodata & \nodata 	& \nodata \\
  2QZ J1225$-$0101 & $I$, $K$ & $>219.4$ & $>225.7$ & $>229.3$ 	& $>244.9$ \\
  2QZ J1247$+$0025 & $I$, $K$ & $221.8^{+39.8}_{-30.1}$ & $230.7^{+37.6}_{-30.7}$ & $235.8^{+35.1}_{-31.9}$ 	& $245.4^{+40.0}_{-32.7}$ \\
 SDSS J1309$-$0015 & $I$, $K$ & $256.4^{+57.4}_{-57.4}$ & $261.2^{+60.2}_{-60.2}$ & $258.5^{+56.0}_{-56.0}$ 	& $278.4^{+64.1}_{-64.1}$ \\
  LBQS 1339$+$0210 & $R$, $K$ & \nodata & \nodata & \nodata 	& \nodata \\
  2QZ J1345$-$0231 & $I$, $K$ & $215.4^{+15.4}_{-13.1}$ & $225.1^{+14.1}_{-13.5}$ & $230.4^{+14.2}_{-13.6}$ 	& $242.9^{+15.2}_{-14.6}$ \\
  2QZ J1438$-$0116 & $I$, $K$ & \nodata & \nodata & \nodata 	& \nodata \\
     PG 1613$+$658 & $R$, $K$ & $320.1^{+43.8}_{-43.4}$ & $326.9^{+41.5}_{-36.1}$ & $327.3^{+40.5}_{-34.5}$ 	& $334.1^{+42.4}_{-37.0}$ \\
 SDSS J1717$+$5932 & $R$, $K$ & $108.1^{+4.4}_{-2.7}$ & $111.0^{+3.6}_{-2.5}$ & $112.7^{+3.7}_{-2.5}$ 	& $115.6^{+3.8}_{-2.6}$ \\
 SDSS J1718$+$5313 & $R$, $K$ & $42.5^{+2.4}_{-2.9}$ & $43.1^{+2.4}_{-3.0}$ & $43.4^{+2.5}_{-2.9}$ 	& $44.5^{+2.5}_{-3.1}$ \\
 SDSS J1720$+$6128 & $I$, $K$ & $95.9^{+2.8}_{-4.5}$ & $98.0^{+6.1}_{-3.0}$ & $107.8^{+10.4}_{-10.4}$ 	& $101.8^{+6.4}_{-3.2}$ \\
 SDSS J1723$+$5400 & $I$, $K$ & $225.0^{+63.4}_{-63.4}$ & $247.8^{+45.7}_{-74.5}$ & $252.0^{+40.3}_{-63.4}$ 	& $265.9^{+49.0}_{-79.9}$ \\
 SDSS J1724$+$6036 & $I$, $K$ & $167.0^{+7.8}_{-7.0}$ & $167.6^{+6.6}_{-6.9}$ & $168.2^{+6.3}_{-7.0}$ 	& $177.4^{+7.0}_{-7.3}$ \\
IRAS F21256$+$0219 & $R$, $K$ & \nodata & \nodata & \nodata 	& \nodata \\
   RX J2138$+$0112 & $I$, $K$ & $173.0^{+9.3}_{-7.7}$ & $178.9^{+7.7}_{-10.3}$ & $182.0^{+7.9}_{-6.2}$ 	& $188.7^{+8.1}_{-10.9}$ \\
   RX J2156$+$1426 & $I$, $K$ & $163.0^{+1.1}_{-1.0}$ & $163.4^{+1.1}_{-1.0}$ & $174.9^{+11.6}_{-11.6}$ 	& $172.6^{+1.1}_{-1.1}$ \\
 SDSS J2326$-$0030 & $I$, $K$ & $188.6^{+55.0}_{-55.0}$ & $212.8^{+65.1}_{-65.1}$ & $226.4^{+70.8}_{-70.8}$ 	& $231.1^{+70.8}_{-70.8}$ \\
\enddata
\tablecomments{}
\tablenotetext{a}{The optical and near-infrared bands between which the dust-reverberation lag was obtained.}
\tablenotetext{b}{The observed dust-reverberation lag.}
\tablenotetext{c}{The dust-reverberation lag for the $K$-band emission in the rest frame.}
 \tablenotetext{d}{The power-law index $\alpha _{\rm OIR}$ assumed for the accretion-disk component between optical and near-infrared wavelengths.}
 \tablenotetext{e}{The observed dust-reverberation lags for the second CCF peak are $314.4^{+11.4}_{-10.1}$, $320.3^{+10.6}_{-12.3}$, $319.8^{+10.5}_{-12.7}$ days for $\alpha_{\rm OIR}=+1/3$, $+0.1$, $0.0$ respectively, and the rest-frame lag for that is $323.9^{+10.7}_{-12.4}$ days.}
 \tablenotetext{f}{The observed dust-reverberation lags for the second CCF peak are $611.6^{+15.8}_{-23.8}$, $614.2^{+16.8}_{-24.6}$, $616.2^{+16.2}_{-25.3}$ days for $\alpha_{\rm OIR}=+1/3$, $+0.1$, $0.0$ respectively, and the rest-frame lag for that is $646.6^{+17.7}_{-25.9}$ days.}

\end{deluxetable*}

\clearpage
\begin{deluxetable*}{lcccccc}
\tablecaption{Host-galaxy Fluxes\label{tab:hostflux}}
\tabletypesize{\footnotesize}
\tablewidth{0pt}
\tablehead{
 \colhead{Object} & \colhead{Band\tablenotemark{\footnotesize a}} & \colhead{$f_{\rm host}$(QSFit)\tablenotemark{\footnotesize b}} &  \colhead{Band2\tablenotemark{\footnotesize c}} & \colhead{$f_{\rm host}$(FVG)\tablenotemark{\footnotesize d}} & \colhead{$f_{\rm host}$(HST)\tablenotemark{\footnotesize e}} & \colhead{$f_{\rm host}$\tablenotemark{\footnotesize f}} \\
 & & ($\mu $Jy) & & ($\mu $Jy) & ($\mu $Jy) & ($\mu $Jy) 
 }
\startdata
 SDSS J0007$-$0054 & $R$ & $140.4^{+1.7}_{-1.7}$ & $V$      & $149.3^{+3.5}_{-4.9}$   & \nodata             &  $156.3^{+6.9}_{-6.9}$ \\
  LBQS 0023$+$0228 & $R$ & \nodata               & \nodata  & \nodata                 & \nodata             &  \nodata \\
 SDSS J0207$-$0048 & $I$ & \nodata               & $R$      & $ 61.4^{+11.0}_{-18.1}$ & \nodata             &  $61.4^{+11.0}_{-18.1}$ \\
 SDSS J0315$+$0012 & $R$ & $131.5^{+2.1}_{-2.1}$ & $V$      & $155.4^{+4.4}_{-5.4}$   & \nodata             &  $146.4^{+6.7}_{-6.7}$ \\
     PG 0844$+$349 & $V$ & \nodata               & $B$      & $1109^{+495}_{-602}$    & $770^{+160}_{-160}$ &  $770^{+160}_{-160}$ \\
 SDSS J0943$-$0043 & $R$ & $ 53.5^{+0.9}_{-0.9}$ & $V$      & $ 46.9^{+8.9}_{-12.4}$  & \nodata             &  $59.5^{+2.7}_{-2.7}$ \\
     PG 0953$+$414 & $R$ & \nodata               & $V$      & $591^{+276}_{-324}$     & $370^{+50}_{-50}$   &  $370^{+50}_{-50}$ \\
 SDSS J0957$-$0023 & $I$ & $121.4^{+1.4}_{-1.4}$ & \nodata  & \nodata                 & \nodata             &  $135.1^{+6.0}_{-6.0}$ \\
  2QZ J1013$+$0028 & $R$ & $ 38.4^{+1.8}_{-1.8}$ & $V$      & $ 50.0^{+4.1}_{-5.6}$   & \nodata             &  $42.7^{+2.5}_{-2.5}$ \\
  LBQS 1022$-$0005 & $R$ & $ 50.5^{+1.1}_{-1.1}$ & $V$      & $-14.6^{+57.8}_{-128.7}$& \nodata             &  $56.2^{+2.6}_{-2.6}$ \\
  LBQS 1026$-$0032 & $R$ & $ 84.8^{+1.9}_{-1.9}$ & $V$      & $100.8^{+7.9}_{-9.3}$   & \nodata             &  $94.4^{+4.5}_{-4.5}$ \\
  2QZ J1032$-$0233 & $I$ & \nodata               & $R$      & $ 31.5^{+34.4}_{-57.1}$ & \nodata             &  $31.5^{+34.4}_{-31.5}$ \\
 SDSS J1044$+$0003 & $I$ & $ 15.8^{+1.3}_{-1.3}$ & $R$      & $ 20.5^{+13.3}_{-21.8}$ & \nodata             &  $17.6^{+1.5}_{-1.5}$ \\
  2QZ J1138$-$0131 & $I$ & \nodata               & $R$      & $ 34.2^{+11.8}_{-64.5}$ & \nodata             &  $34.2^{+11.8}_{-34.2}$ \\
  2QZ J1220$-$0119 & $I$ & \nodata               & \nodata  & \nodata                 & \nodata             &  \nodata \\
  2QZ J1225$-$0101 & $I$ & \nodata               & $R$      & $  5.5^{+5.6}_{-6.9}$   & \nodata             &  $5.5^{+5.6}_{-5.5}$ \\
  2QZ J1247$+$0025 & $I$ & \nodata               & $R$      & $ 17.1^{+9.2}_{-16.8}$  & \nodata             &  $17.1^{+9.2}_{-16.8}$ \\
 SDSS J1309$-$0015 & $I$ & $ 28.0^{+2.1}_{-2.1}$ & \nodata  & \nodata                 & \nodata             &  $31.1^{+2.5}_{-2.5}$ \\
  LBQS 1339$+$0210 & $R$ & $ 84.2^{+1.5}_{-1.5}$ & $V$      & $ 76.6^{+20.0}_{-38.1}$ & \nodata             &  $93.7^{+4.3}_{-4.3}$ \\
  2QZ J1345$-$0231 & $I$ & $ 49.8^{+1.9}_{-1.9}$ & $R$      & $ 48.6^{+3.9}_{-4.6}$   & \nodata             &  $55.4^{+3.1}_{-3.1}$ \\
  2QZ J1438$-$0116 & $I$ & \nodata               & \nodata  & \nodata                 & \nodata             &  \nodata \\
     PG 1613$+$658 & $R$ & \nodata               & $V$      & $-369^{+337}_{-389}$    & $1530^{+70}_{-70}$  &  $1530^{+70}_{-70}$ \\
 SDSS J1717$+$5932 & $R$ & \nodata               & $V$      & $114.0^{+1.5}_{-1.6}$   & $107^{+5}_{-5}$     &  $107^{+5}_{-5}$ \\
 SDSS J1718$+$5313 & $R$ & $ 80.0^{+0.3}_{-0.3}$ & $V$      & $ 86.6^{+0.7}_{-0.8}$ \tablenotemark{\footnotesize g} & \nodata             &  $80.0^{+0.3}_{-0.3}$ \\
 SDSS J1720$+$6128 & $I$ & $242.8^{+2.2}_{-2.2}$ & $R$      & $377.4^{+6.5}_{-7.4}$ \tablenotemark{\footnotesize g} & \nodata             &  $270.3^{+11.9}_{-11.9}$ \\
 SDSS J1723$+$5400 & $I$ & $ 28.2^{+1.4}_{-1.4}$ & $R$      & $ 59.2^{+6.2}_{-8.6}$   & \nodata             &  $31.4^{+1.9}_{-1.9}$ \\
 SDSS J1724$+$6036 & $I$ & $ 31.8^{+1.8}_{-1.8}$ & $R$      & $ 37.1^{+16.1}_{-18.8}$ & \nodata             &  $35.4^{+2.3}_{-2.3}$ \\
IRAS F21256$+$0219 & $R$ & \nodata               & \nodata  & \nodata                 & \nodata             &  \nodata \\
   RX J2138$+$0112 & $I$ & $124.3^{+2.1}_{-2.1}$ & $R$      & $132.2^{+9.5}_{-10.3}$  & \nodata             &  $138.4^{+6.3}_{-6.3}$ \\
   RX J2156$+$1426 & $I$ & \nodata               & $R$      & $ 76.5^{+2.7}_{-2.9}$   & \nodata             &  $76.5^{+2.7}_{-2.9}$ \\
 SDSS J2326$-$0030 & $I$ & $ 37.4^{+2.7}_{-2.7}$ & $R$      & $ 21.6^{+9.2}_{-12.5}$  & \nodata             &  $41.6^{+3.2}_{-3.2}$ \\
\enddata
\tablecomments{}
\tablenotetext{a}{The monitoring optical band for the dust reverberation, and the band for the host-galaxy fluxes presented here}
\tablenotetext{b}{The host-galaxy flux estimated by the spectral decomposition method}
\tablenotetext{c}{The other optical band for the FVG analysis}
\tablenotetext{d}{The host-galaxy flux estimated by the FVG method}
\tablenotetext{e}{The host-galaxy flux estimated by the image-decomposition method}
\tablenotetext{f}{The host-galaxy flux adopted}
 \tablenotetext{g}{The best estimate of the host-galaxy flux exceeds the minimum flux
 during the monitoring observation.}
\end{deluxetable*}

\clearpage
\begin{deluxetable*}{lccc}
\tablecaption{Optical AGN Fluxes and Luminosities\label{tab:optfL}}
\tabletypesize{\footnotesize}
\tablewidth{0pt}
\tablehead{
 \colhead{Object} & \colhead{Band\tablenotemark{\footnotesize a}} & \colhead{$f_{\rm AGN}$\tablenotemark{\footnotesize b}} & \colhead{$\log (\lambda L_{\lambda ,\ {\rm AGN}}(V)/{\rm erg\ s}^{-1})$\tablenotemark{\footnotesize c}} \\
 & & ($\mu $Jy) &
 }
\startdata
 SDSS J0007$-$0054 & $R$ &  $ 16.3^{+ 7.0}_{- 7.0}$ &  $42.602^{+0.154}_{-0.241}$ \\
  LBQS 0023$+$0228 & $R$ &  \nodata                 &  \nodata                    \\
 SDSS J0207$-$0048 & $I$ &  $ 54.3^{+18.1}_{-11.1}$ &  $43.975^{+0.125}_{-0.099}$ \\
 SDSS J0315$+$0012 & $R$ &  $ 54.5^{+ 6.7}_{- 6.7}$ &  $43.454^{+0.050}_{-0.057}$ \\
     PG 0844$+$349 & $V$ &  $ 5878^{+161}_{-161}$   &  $44.454^{+0.012}_{-0.012}$ \\
 SDSS J0943$-$0043 & $R$ &  $ 49.6^{+ 2.7}_{- 2.7}$ &  $43.671^{+0.023}_{-0.025}$ \\
     PG 0953$+$414 & $R$ &  $ 3845^{+52}_{-52}$     &  $45.423^{+0.006}_{-0.006}$ \\
 SDSS J0957$-$0023 & $I$ &  $523.4^{+ 6.0}_{- 6.0}$ &  $45.408^{+0.005}_{-0.005}$ \\
  2QZ J1013$+$0028 & $R$ &  $ 35.6^{+ 2.6}_{- 2.6}$ &  $43.712^{+0.030}_{-0.032}$ \\
  LBQS 1022$-$0005 & $R$ &  $137.7^{+ 2.7}_{- 2.7}$ &  $44.270^{+0.008}_{-0.008}$ \\
  LBQS 1026$-$0032 & $R$ &  $102.3^{+ 4.5}_{- 4.5}$ &  $43.944^{+0.019}_{-0.020}$ \\
  2QZ J1032$-$0233 & $I$ &  $165.9^{+31.5}_{-34.4}$ &  $44.863^{+0.075}_{-0.101}$ \\
 SDSS J1044$+$0003 & $I$ &  $ 69.7^{+ 1.5}_{- 1.5}$ &  $44.252^{+0.010}_{-0.010}$ \\
  2QZ J1138$-$0131 & $I$ &  $ 30.8^{+34.2}_{-11.8}$ &  $43.971^{+0.324}_{-0.210}$ \\
  2QZ J1220$-$0119 & $I$ &  \nodata                 &  \nodata                    \\
  2QZ J1225$-$0101 & $I$ &  $ 60.6^{+ 5.5}_{- 5.6}$ &  $44.440^{+0.038}_{-0.042}$ \\
  2QZ J1247$+$0025 & $I$ &  $ 40.2^{+16.8}_{- 9.2}$ &  $43.944^{+0.152}_{-0.113}$ \\
 SDSS J1309$-$0015 & $I$ &  $ 62.3^{+ 2.5}_{- 2.5}$ &  $44.161^{+0.017}_{-0.018}$ \\
  LBQS 1339$+$0210 & $R$ &  $ 67.2^{+ 4.3}_{- 4.3}$ &  $43.812^{+0.027}_{-0.029}$ \\
  2QZ J1345$-$0231 & $I$ &  $ 41.4^{+ 3.1}_{- 3.1}$ &  $44.195^{+0.031}_{-0.033}$ \\
  2QZ J1438$-$0116 & $I$ &  \nodata                 &  \nodata                    \\
     PG 1613$+$658 & $R$ &  $ 3160^{+74}_{-74}$     &  $44.793^{+0.010}_{-0.010}$ \\
 SDSS J1717$+$5932 & $R$ &  $ 39.4^{+ 5.0}_{- 5.0}$ &  $43.489^{+0.052}_{-0.059}$ \\
 SDSS J1718$+$5313 & $R$ &  $ 10.0^{+ 0.4}_{- 0.4}$ &  $42.653^{+0.017}_{-0.017}$ \\
 SDSS J1720$+$6128 & $R$ &  $110.4^{+11.1}_{-11.1}$ &  $43.890^{+0.042}_{-0.046}$ \\
 SDSS J1723$+$5400 & $I$ &  $ 69.8^{+ 2.0}_{- 2.0}$ &  $44.330^{+0.012}_{-0.013}$ \\
 SDSS J1724$+$6036 & $I$ &  $217.2^{+ 2.5}_{- 2.5}$ &  $44.579^{+0.005}_{-0.005}$ \\
IRAS F21256$+$0219 & $R$ &  \nodata                 &  \nodata                    \\
   RX J2138$+$0112 & $I$ &  $256.0^{+ 6.3}_{- 6.3}$ &  $44.576^{+0.011}_{-0.011}$ \\
   RX J2156$+$1426 & $I$ &  $ 57.3^{+ 2.9}_{- 2.7}$ &  $43.958^{+0.021}_{-0.021}$ \\
 SDSS J2326$-$0030 & $I$ &  $ 49.5^{+ 3.2}_{- 3.2}$ &  $44.359^{+0.027}_{-0.029}$ \\
\enddata
\tablecomments{}
\tablenotetext{a}{The optical band for the $f_{\rm AGN}$}
\tablenotetext{b}{The optical flux of the AGN}
\tablenotetext{c}{The $V$-band luminosity of the AGN}
\end{deluxetable*}

\clearpage
\begin{deluxetable*}{lcccccc}
\tablecaption{Dust-reverberation Lag and Optical AGN Luminosities of the Seyfert Galaxies from Koshida et al. (2014)\label{tab:Seyfert}}
\tabletypesize{\footnotesize}
 \tablewidth{0pt}
\tablehead{
 \colhead{Object} & \colhead{Redshift\tablenotemark{\footnotesize a}} & \colhead{Redshift\tablenotemark{\footnotesize b}} & \colhead{Lag Time\tablenotemark{\footnotesize c}} & \colhead{Lag Time\tablenotemark{\footnotesize d}} & \colhead{Flux ($V$)\tablenotemark{\footnotesize e}} & \colhead{$\log (\lambda L_{\lambda ,\ {\rm AGN}}(V)/{\rm erg\ s}^{-1})$\tablenotemark{\footnotesize f}} \\
 & (km s$^{-1}$) & (km s$^{-1}$) & (days) & (days) & (mJy) &  
 }
 \startdata
          Mrk 335 & $0.025785$ & $0.025341$ & $166.7 \pm  5.9$  & $167.5 \pm  6.0$  & $ 5.899\pm 0.124$ & $43.631^{+0.009}_{-0.009}$ \\
          Mrk 590 & $0.026385$ & $0.025638$ & $ 33.32\pm  4.14$ & $ 33.48\pm  4.16$ & $ 0.965\pm 0.074$ & $42.854^{+0.032}_{-0.035}$ \\
IRAS 03450$+$0055 & $0.031000$ & $0.030261$ & $156.6 \pm  5.9$  & $157.4 \pm  5.9$  & $ 7.688\pm 0.155$ & $43.901^{+0.009}_{-0.009}$ \\
          Akn 120 & $0.032713$ & $0.032256$ & $137.5 \pm 17.4$  & $138.3 \pm 17.5$  & $15.12 \pm 0.29 $ & $44.252^{+0.008}_{-0.008}$ \\
 MCG $+08-11-011$ & $0.020484$ & $0.020551$ & $ 72.42\pm  1.61$ & $ 72.68\pm  1.62$ & $ 5.726\pm 0.319$ & $43.434^{+0.024}_{-0.025}$ \\
           Mrk 79 & $0.022189$ & $0.022662$ & $ 67.46\pm  4.76$ & $ 67.73\pm  4.78$ & $ 3.998\pm 0.135$ & $43.364^{+0.014}_{-0.015}$ \\
          Mrk 110 & $0.035291$ & $0.036008$ & $115.8 \pm  6.3$  & $116.6 \pm  6.3$  & $ 2.941\pm 0.103$ & $43.638^{+0.015}_{-0.015}$ \\
         NGC 3227 & $0.003859$ & $0.004953$ & $ 14.36\pm  0.70$ & $ 14.37\pm  0.70$ & $ 5.879\pm 0.323$ & $42.203^{+0.023}_{-0.025}$ \\
         NGC 3516 & $0.008836$ & $0.010060$ & $ 72.62\pm  4.54$ & $ 72.74\pm  4.55$ & $ 3.798\pm 0.320$ & $42.630^{+0.035}_{-0.038}$ \\
          Mrk 744 & $0.008933$ & $0.010621$ & $ 19.91\pm  2.20$ & $ 19.94\pm  2.20$ & $ 0.506\pm 0.028$ & $41.804^{+0.023}_{-0.024}$ \\
         NGC 4051 & $0.002336$ & $0.003146$ & $ 16.29\pm  0.57$ & $ 16.30\pm  0.57$ & $ 4.390\pm 0.379$ & $41.682^{+0.036}_{-0.039}$ \\
         NGC 4151 & $0.003319$ & $0.004857$ & $ 46.08\pm  0.44$ & $ 46.11\pm  0.44$ & $16.52 \pm 1.34 $ & $42.635^{+0.034}_{-0.037}$ \\
         NGC 4593 & $0.009000$ & $0.010057$ & $ 41.75\pm  0.90$ & $ 41.82\pm  0.90$ & $ 3.175\pm 0.120$ & $42.553^{+0.016}_{-0.017}$ \\
         NGC 5548 & $0.017175$ & $0.019267$ & $ 61.02\pm  0.30$ & $ 61.21\pm  0.30$ & $ 1.849\pm 0.058$ & $42.887^{+0.013}_{-0.014}$ \\
          Mrk 817 & $0.031455$ & $0.032769$ & $ 92.04\pm  8.81$ & $ 92.55\pm  8.86$ & $ 4.405\pm 0.076$ & $43.731^{+0.007}_{-0.008}$ \\
          Mrk 509 & $0.034397$ & $0.035018$ & $120.6 \pm  1.6$  & $121.3 \pm  1.6$  & $12.12 \pm 0.29 $ & $44.228^{+0.010}_{-0.010}$ \\
         NGC 7469 & $0.016317$ & $0.016315$ & $ 85.04\pm  0.43$ & $ 85.29\pm  0.43$ & $ 6.472\pm 0.375$ & $43.286^{+0.024}_{-0.026}$ \\
\enddata
\tablecomments{}
\tablenotetext{a}{The heliocentric redshift from the NED.}
\tablenotetext{b}{ The local flow-corrected redshift of the Virgo infall
 $+$ Great Attractor $+$ Shapley supercluster from the NED.}
\tablenotetext{c}{The observed dust-reverberation lag where $\alpha _{VK}=+0.1$ was assumed for the near-infrared continuum emission of the accretion disk}
\tablenotetext{d}{The dust-reverberation lag for the $K$-band emission in the rest frame.}
\tablenotetext{e}{The $V$-band flux of the AGN; Galactic extinction has been corrected for and the flux from the host galaxy has been subtracted.}
\tablenotetext{f}{The $V$-band luminosity of the AGN}
\end{deluxetable*}

\clearpage
\begin{deluxetable*}{ccrcccc}
\tablecaption{Results of Linear Regression of Dust-Reverberation Radius--Luminosity Correlation\label{tab:rL}}
\tabletypesize{\footnotesize}
 \tablewidth{0pt}
\tablehead{
 \colhead{$L$ (erg s$^{-1}$)} & \colhead{$\log (L_{0}/{\rm erg\ s}^{-1})$} & \colhead{$N$\tablenotemark{\footnotesize a}} & \colhead{$\alpha $} & \colhead{$\beta $\tablenotemark{\footnotesize b}} & \colhead{$\sigma _{\rm add}$} & \colhead{Method\tablenotemark{\footnotesize c}}
 }
\startdata
$\lambda L_{\lambda} (V)$ & 43.7 & 36 & $-1.021 \pm 0.025$ & 0.5               & 0.14 & 1\\
                          &      &    & $-1.021 \pm 0.023$ & $0.424 \pm 0.026$ & 0.12 & 1\\
                          &      &    & $-1.021 \pm 0.022$ & $0.423 \pm 0.026$ & 0.12 & 2\\
                          &      &    & $-1.024^{+0.024}_{-0.023}$ & $0.427^{+0.028}_{-0.027}$ & 0.13 & 3\\
$L_{\rm BAT}$             & 43.5 & 15 & $-1.27  \pm 0.05$  & 0.5               & 0.19 & 1\\
                          &      &    & $-1.27  \pm 0.05$  & $0.44 \pm 0.07$   & 0.19 & 1\\
                          &      &    & $-1.27  \pm 0.05$  & $0.44 \pm 0.07$   & 0.19 & 2\\
                          &      &    & $-1.28^{+0.06}_{-0.06}$ & $0.44^{+0.08}_{-0.08}$ & 0.21 & 3\\
$L_{[\rm O_{IV}]}$        & 41.0 & 15 & $-1.22  \pm 0.06$  & 0.5               & 0.23 & 1\\
                          &      &    & $-1.22  \pm 0.06$  & $0.52 \pm 0.09$   & 0.24 & 1\\
                          &      &    & $-1.22  \pm 0.06$  & $0.52 \pm 0.08$   & 0.23 & 2\\
                          &      &    & $-1.23^{+0.07}_{-0.07}$ & $0.53^{+0.10}_{-0.10}$ & 0.27 & 3\\
$\lambda L_{\lambda} (12\mu {\rm m})$ & 43.5 & 8 & $-1.33 \pm 0.04$ & 0.5             & 0.09 & 1\\
                                      &      &   & $-1.33 \pm 0.05$ & $0.50 \pm 0.07$ & 0.10 & 1\\
                          &      &               & $-1.33 \pm 0.05$ & $0.50 \pm 0.07$ & 0.10 & 2\\
                          &      &               & $-1.33^{+0.07}_{-0.07}$ & $0.50^{+0.11}_{-0.10}$ & 0.16 & 3\\
\enddata
\tablecomments{The fitted model is $\log r = \alpha + \beta \, (\log L - \log L_{0})$}
\tablenotetext{a}{The number of the data pair for the fitting.}
\tablenotetext{b}{The parameter was fixed when $\beta = 0.5$.}
\tablenotetext{c}{(1) \citet{1980AJ.....85..177J,1981AJ.....86..149J}, (2) \citet{1992nrfa.book.....P}, \citet{2010MNRAS.409.1330W}, (3) \citet{2007ApJ...665.1489K}}
\end{deluxetable*}

\clearpage
\begin{deluxetable*}{lcccc}
\tablecaption{Isotropic Luminosity Indicators for the AGNs\label{tab:isoL}}
\tabletypesize{\footnotesize}
 \tablewidth{0pt}
\tablehead{
 \colhead{Object} & \colhead{$\log (L_{\rm BAT}/{\rm erg\ s}^{-1})$\tablenotemark{\footnotesize a}} & \colhead{$\log (L_{\rm [O_{\rm IV}]}/{\rm erg\ s}^{-1})$\tablenotemark{\footnotesize b}} & \colhead{References\tablenotemark{\footnotesize c}} & \colhead{$\log (\lambda L_{\lambda }(12\mu {\rm m})/{\rm erg\ s}^{-1})$\tablenotemark{\footnotesize d}} 
 }
 \startdata
          Mrk 335 &  $43.36$ &  $41.01$ & 1 & \nodata         \\
          Mrk 590 &  $43.35$ &  $40.66$ & 2 & $43.52\pm 0.04$ \\
IRAS 03450$+$0055 &  \nodata &  $40.7$  & 3 & \nodata         \\
          Akn 120 &  $44.17$ &  $40.95$ & 4 & $44.12\pm 0.02$ \\
 MCG $+08-11-011$ &  $44.06$ &  \nodata & \nodata & \nodata   \\
           Mrk 79 &  $43.66$ &  $41.72$ & 1 & \nodata         \\
          Mrk 110 &  $44.17$ &  $41.06$ & 5 & \nodata         \\
         NGC 3227 &  $42.55$ &  $40.23$ & 1 & $42.40\pm 0.11$ \\
         NGC 3516 &  $43.29$ &  $40.95$ & 1 & \nodata         \\
          Mrk 744 &  $42.45$ &  $40.48$ & 1 & \nodata         \\
         NGC 4051 &  $41.65$ &  $39.46$ & 1 & $42.36\pm 0.04$ \\
         NGC 4151 &  $43.10$ &  $40.66$ & 1 & $43.19\pm 0.08$ \\
         NGC 4593 &  $43.20$ &  $40.34$ & 1 & $43.07\pm 0.07$ \\
         NGC 5548 &  $43.66$ &  $40.96$ & 1 & $43.37\pm 0.27$ \\
          Mrk 817 &  $43.73$ &  $41.1$  & 6 & \nodata         \\
          Mrk 509 &  $44.37$ &  $41.85$ & 1 & $44.21\pm 0.05$ \\
         NGC 7469 &  $43.55$ &  $41.30$ & 1 & $43.82\pm 0.05$ \\
    PG 0844$+$349 &  \nodata &  $41.16$ & 2 & $43.93\pm 0.17$ \\
    PG 0953$+$414 &  \nodata &  $42.27$ & 2 & \nodata \\
    PG 1613$+$658 &  $44.70$ &  $42.37$ & 2 & \nodata \\
 \enddata
\tablecomments{}
\tablenotetext{a}{The hard X-ray luminosity at 14--195 keV}
\tablenotetext{b}{The luminosity of [O {\scriptsize ${\rm IV}$}] 25.89 $\mu$m emission lines}
\tablenotetext{c}{References for $L_{\rm [O_{\rm IV}]}$ : 1-- \citet{2010ApJ...725.2381L}, in which many data come from \citet{2008ApJ...682...94M} and \citet{2009ApJ...698..623D}, 2-- \citet{2010ApJ...723..409G}, 3-- \citet{2010ApJ...709.1257T}, 4-- \citet{2010ApJ...716.1151W}, 5-- \citet{2018MNRAS.478.4068L}, 6-- \citet{2007ApJ...671..124D}}
\tablenotetext{d}{The nuclear luminosity of mid-infrared continuum $\lambda L_{\lambda }$ at $\lambda=12$ $\mu$m}
\end{deluxetable*}

\clearpage
\begin{figure}
\plotone{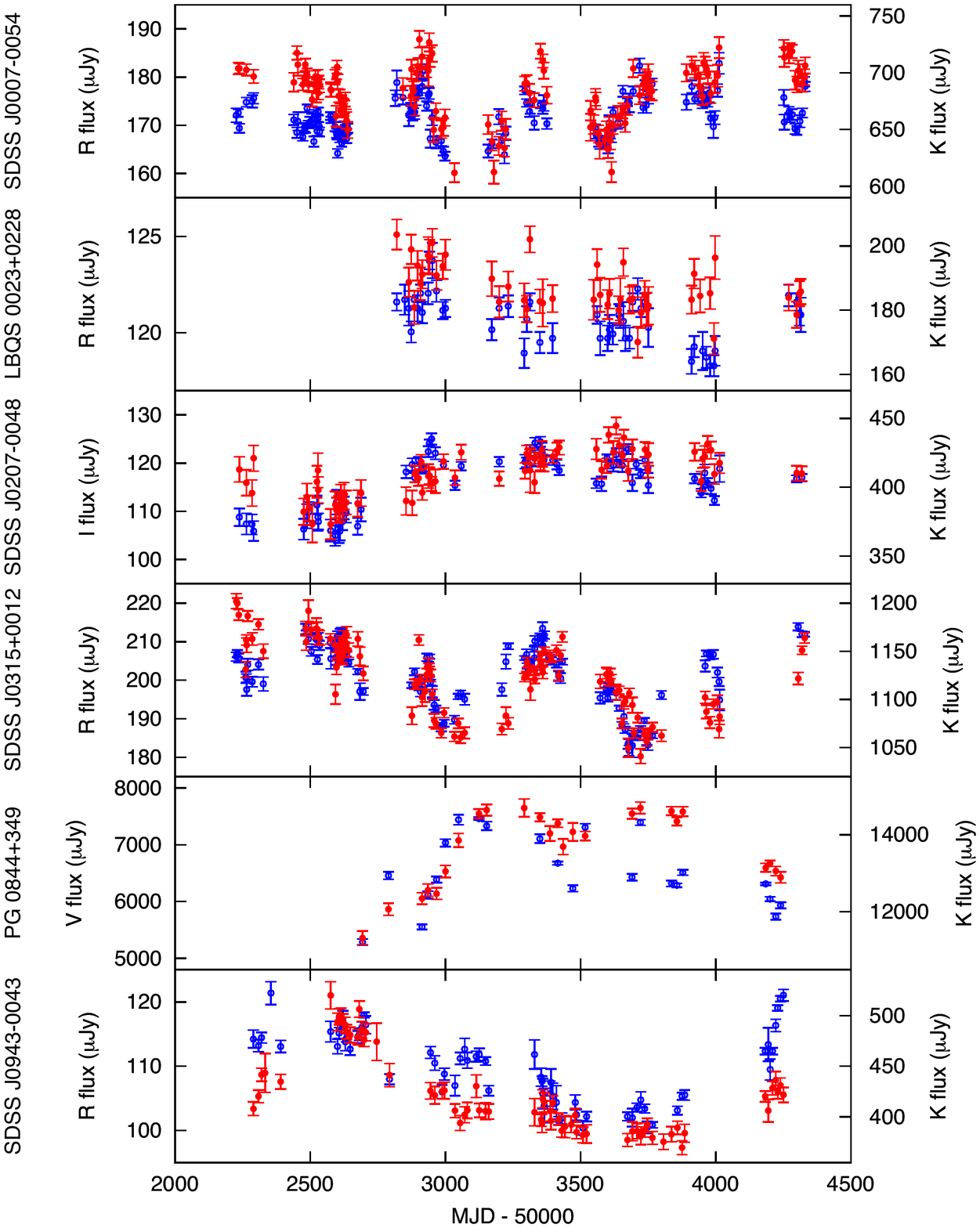}
 \caption{Optical and $K$-band light curves of the target quasars. Blue open circles represent the optical fluxes, and red filled circles represent the $K$-band fluxes. Correction for the Galactic extinction has been applied to the light curves, and the fluxes from the host galaxy have not been subtracted.
 \label{fig:lc1}}
\end{figure}

\clearpage
\begin{figure}
\figurenum{1}
\plotone{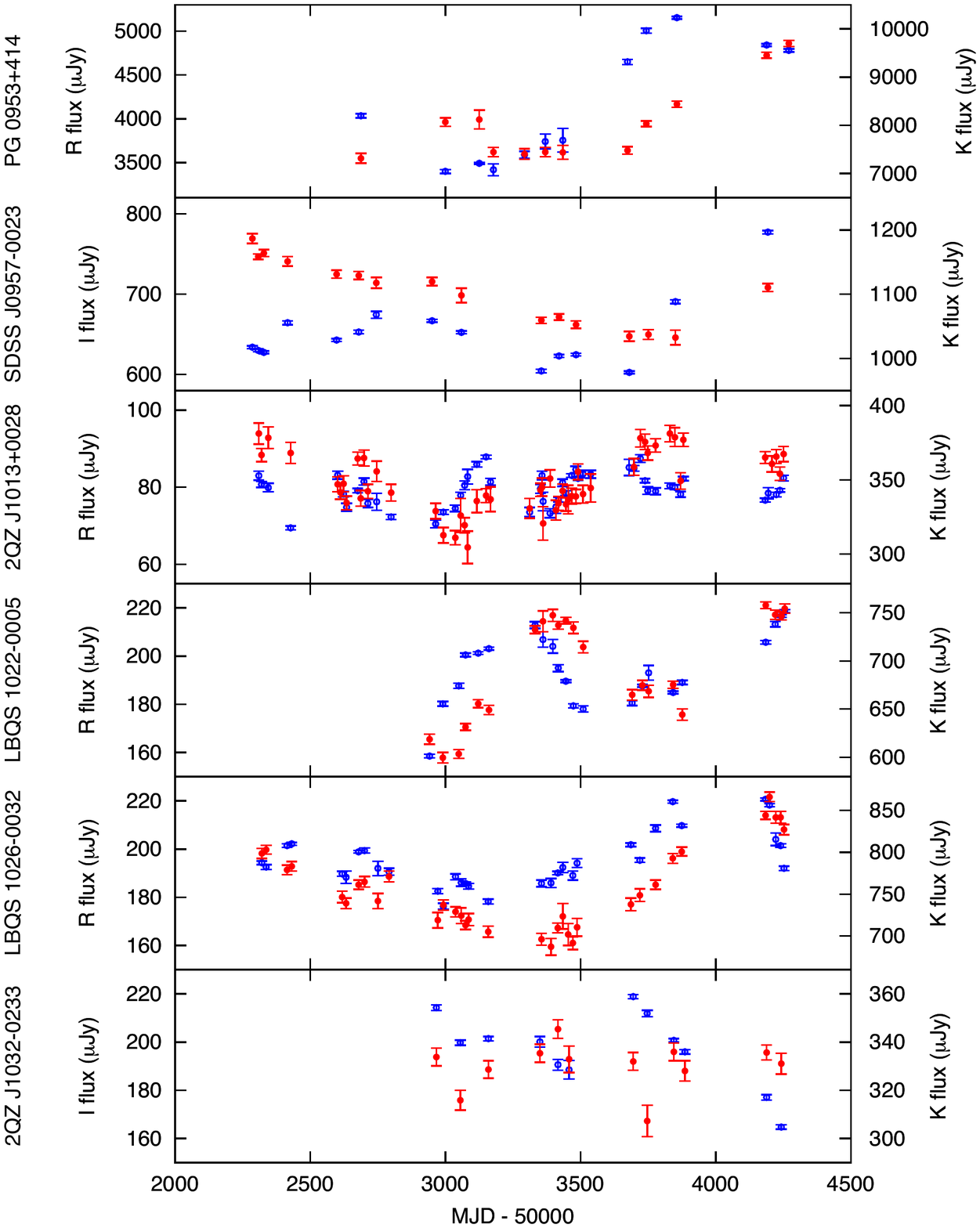}
 \caption{(Continued.)
 \label{fig:lc2}}
\end{figure}

\clearpage
\begin{figure}
\figurenum{1}
\plotone{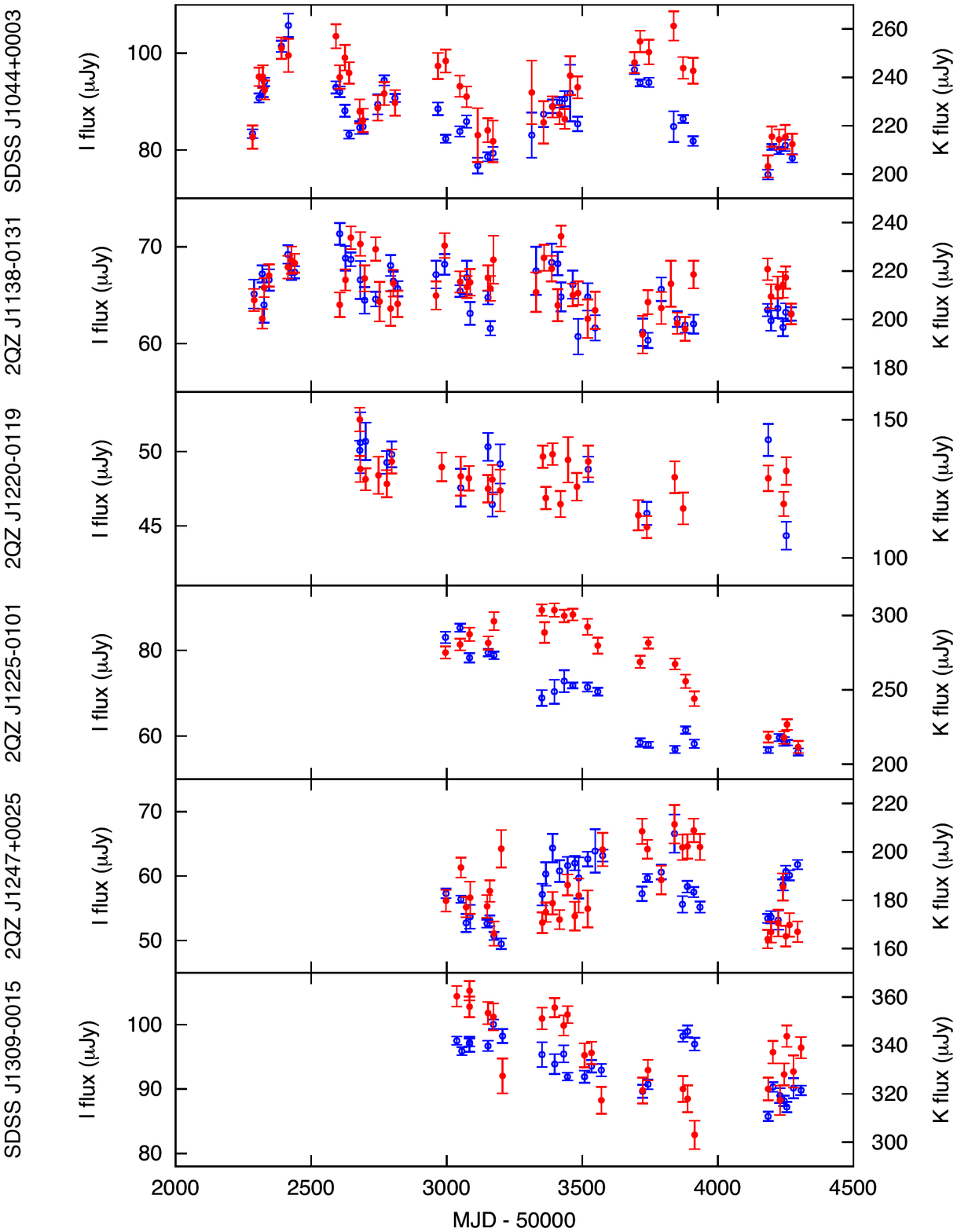}
 \caption{(Continued.)
 \label{fig:lc3}}
\end{figure}

\clearpage
\begin{figure}
\figurenum{1}
\plotone{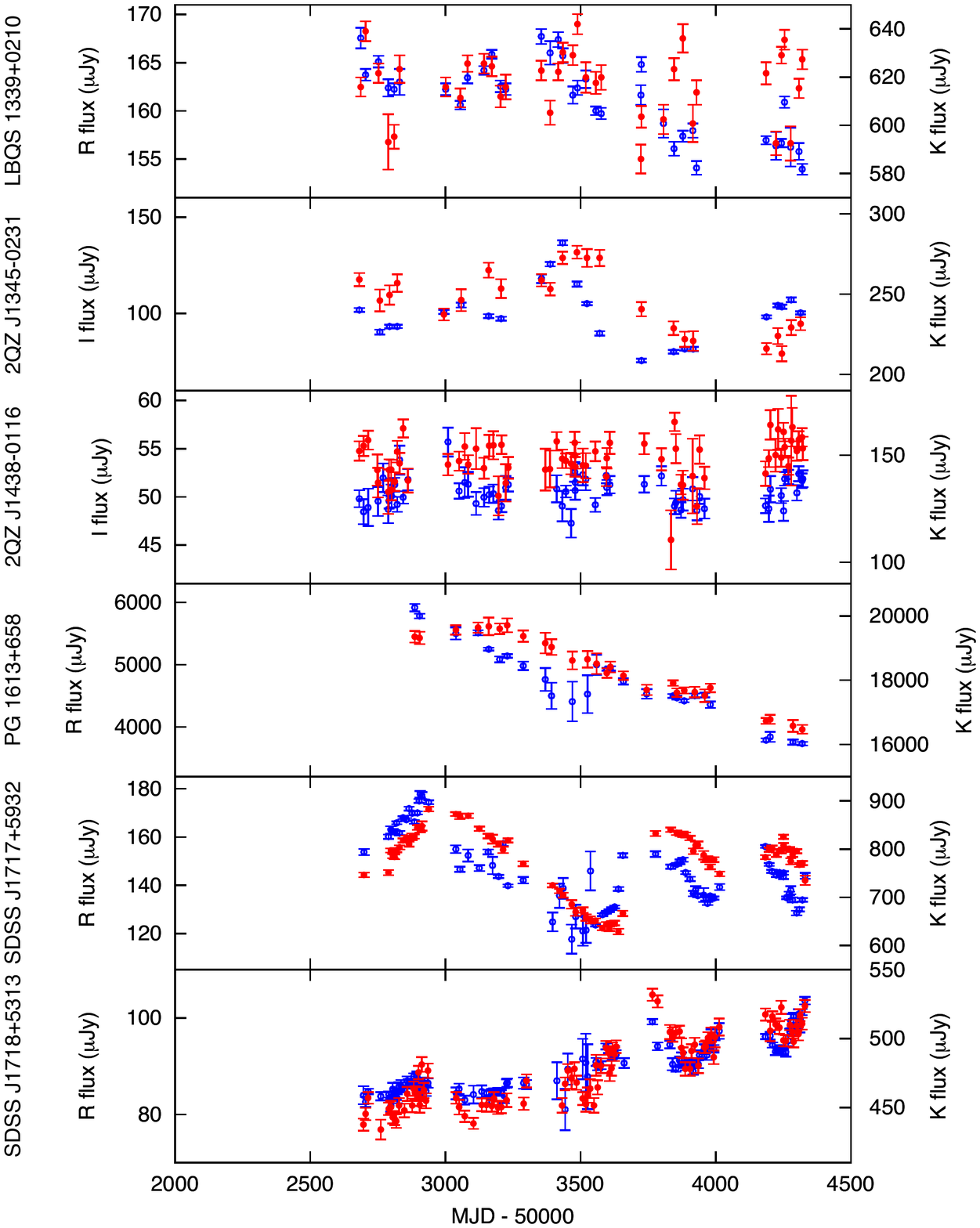}
 \caption{(Continued.)
 \label{fig:lc4}}
\end{figure}

\clearpage
\begin{figure}
\figurenum{1}
\plotone{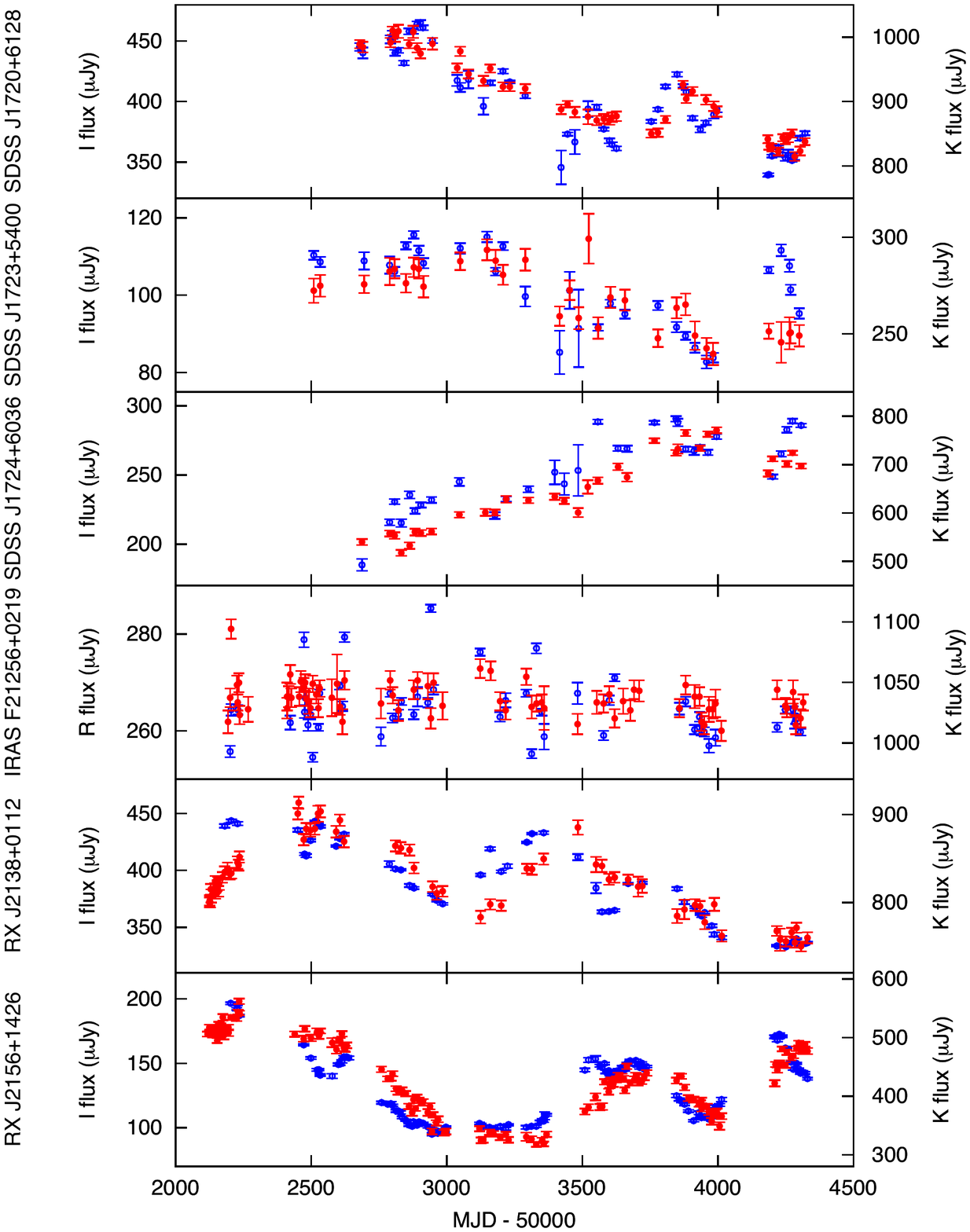}
 \caption{(Continued.)
 \label{fig:lc5}}
\end{figure}

\clearpage
\begin{figure}
\figurenum{1}
\plotone{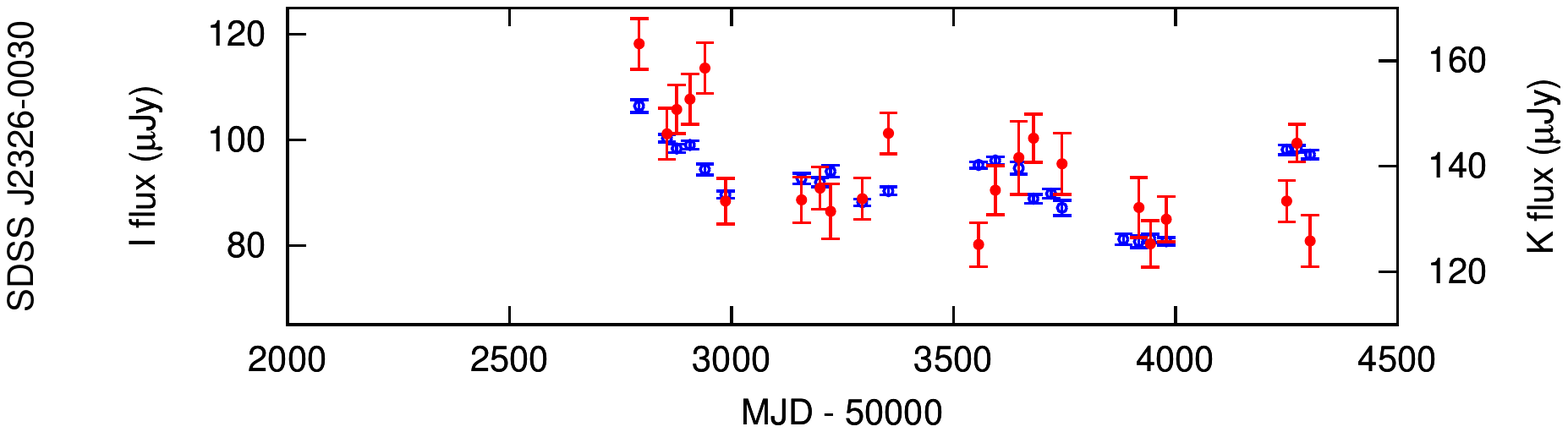}
 \caption{(Continued.)
 \label{fig:lc6}}
\end{figure}

\clearpage
\begin{figure}
\plotone{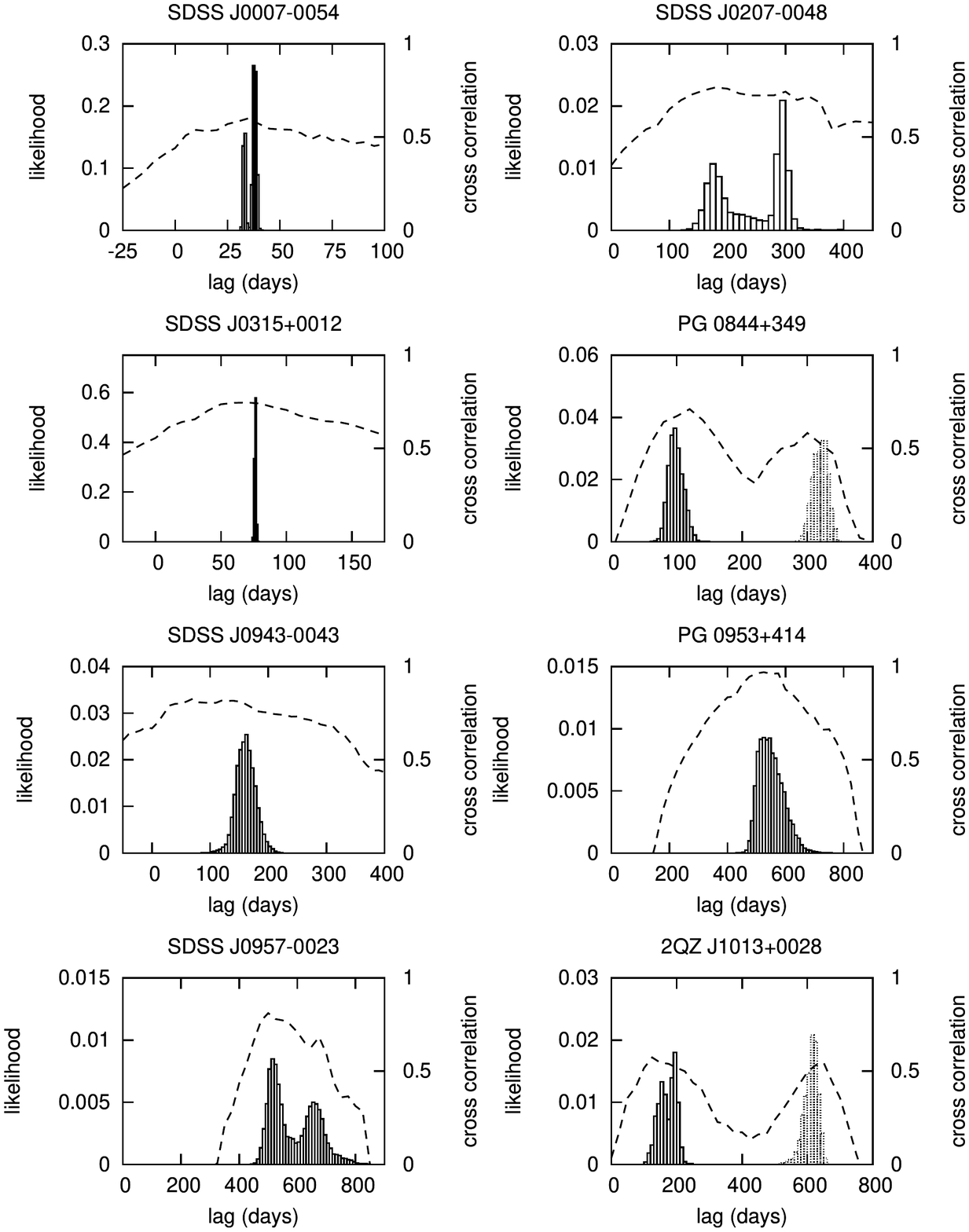}
 \caption{Likelihood distributions of the lag time and cross-correlation functions (CCFs) between the optical and $K$-band flux variations of the target quasars. Histograms represent the likelihood distributions estimated by JAVELIN, and dashed lines represent the CCF. Dotted histograms represent the likelihood distribution for the secondary CCF peak.
 \label{fig:lag1}}
\end{figure}

\clearpage
\begin{figure}
\figurenum{2}
\plotone{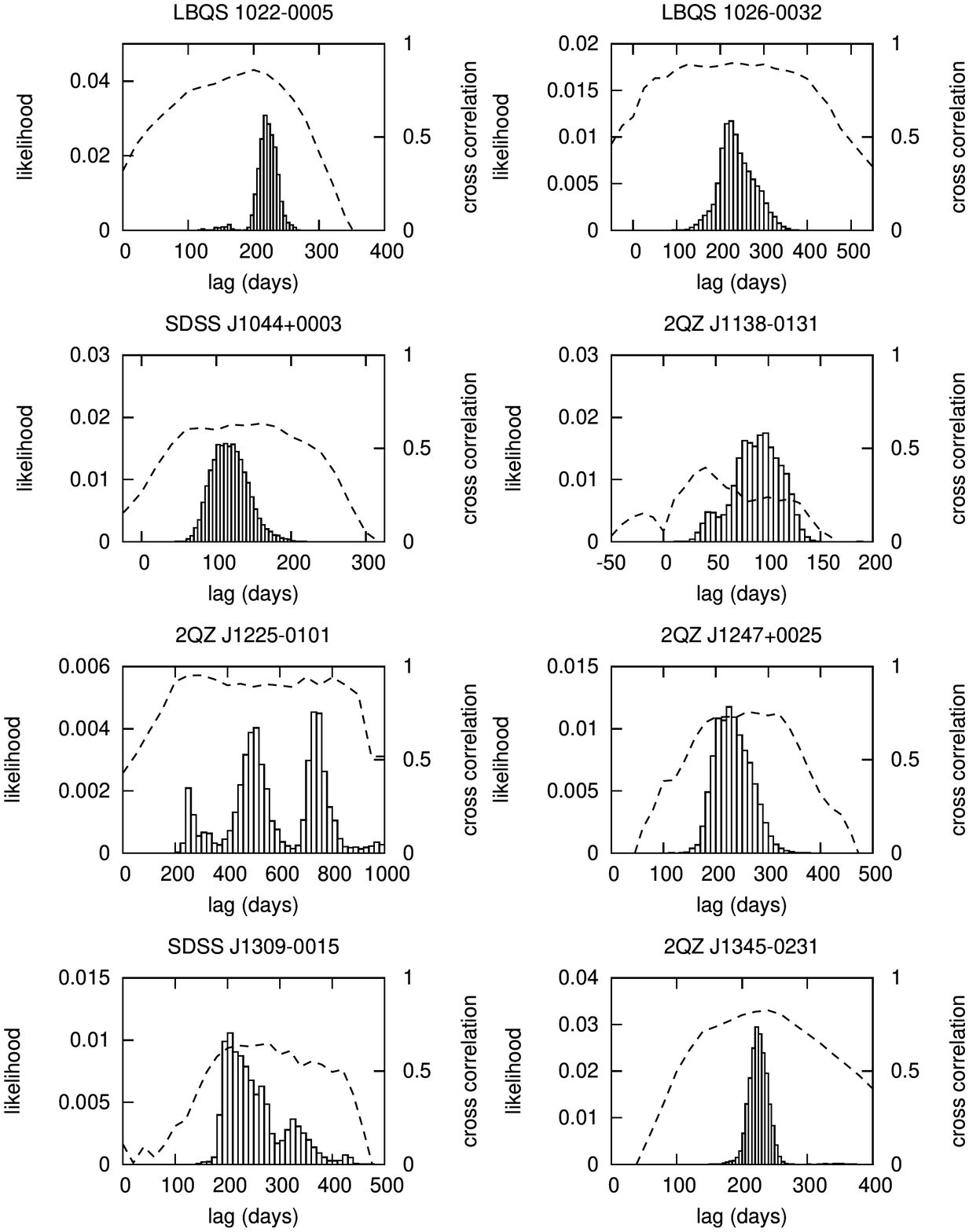}
 \caption{(Continued.)
 \label{fig:lag2}}
\end{figure}

\clearpage
\begin{figure}
\figurenum{2}
\plotone{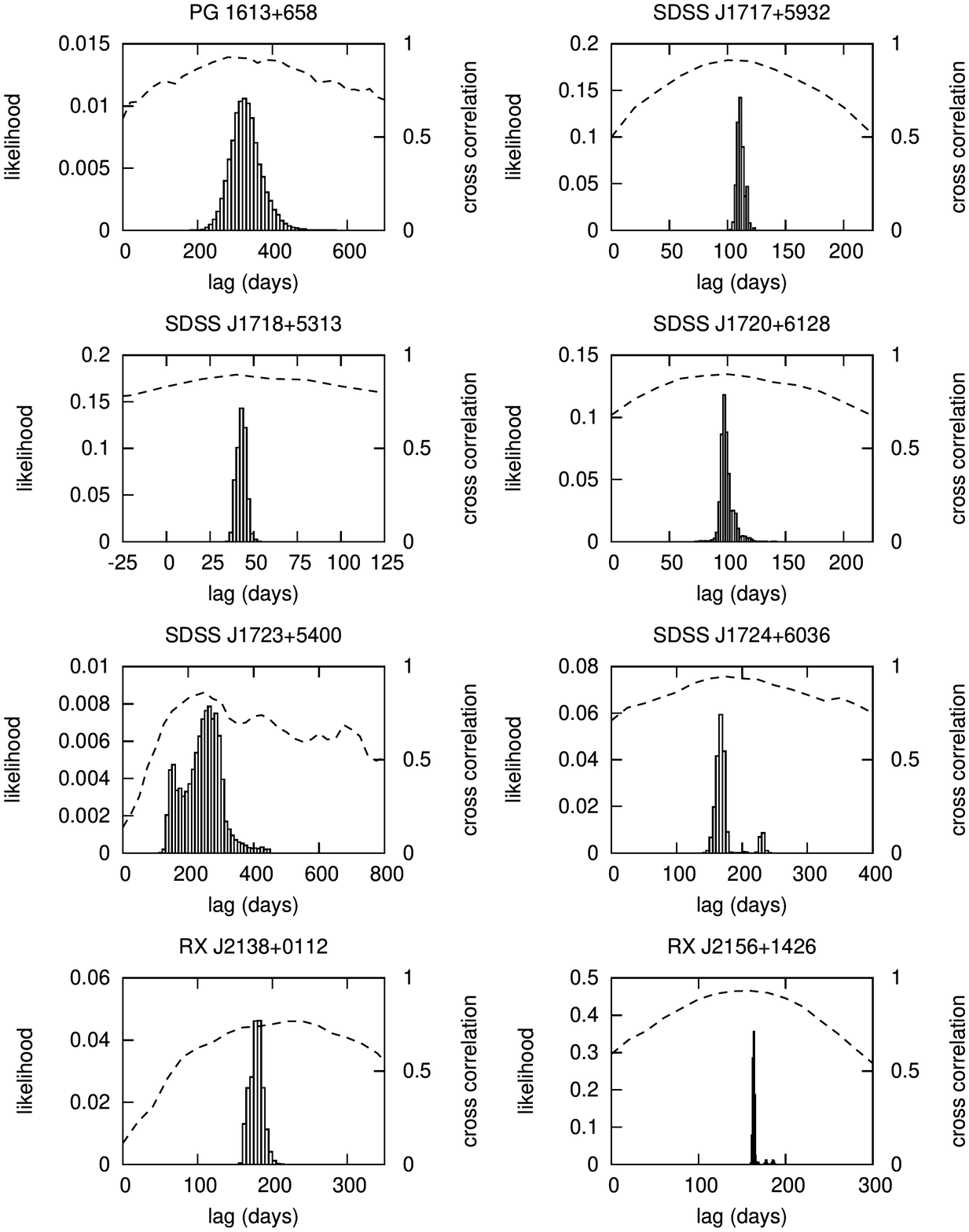}
 \caption{(Continued.)
 \label{fig:lag3}}
\end{figure}

\clearpage
\begin{figure}
\figurenum{2}
\plotone{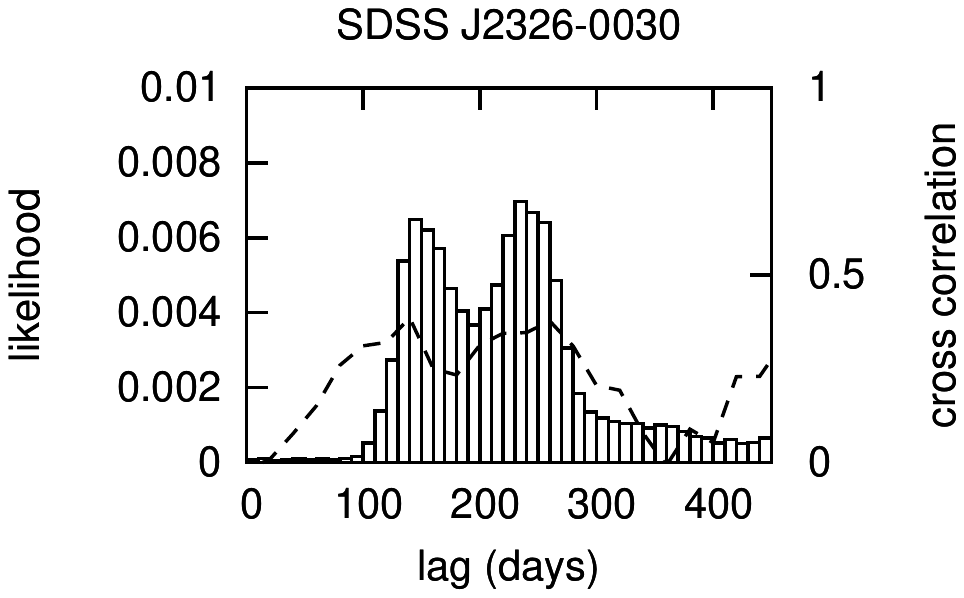}
 \caption{(Continued.)
 \label{fig:lag4}}
\end{figure}

\clearpage
\begin{figure}
\plotone{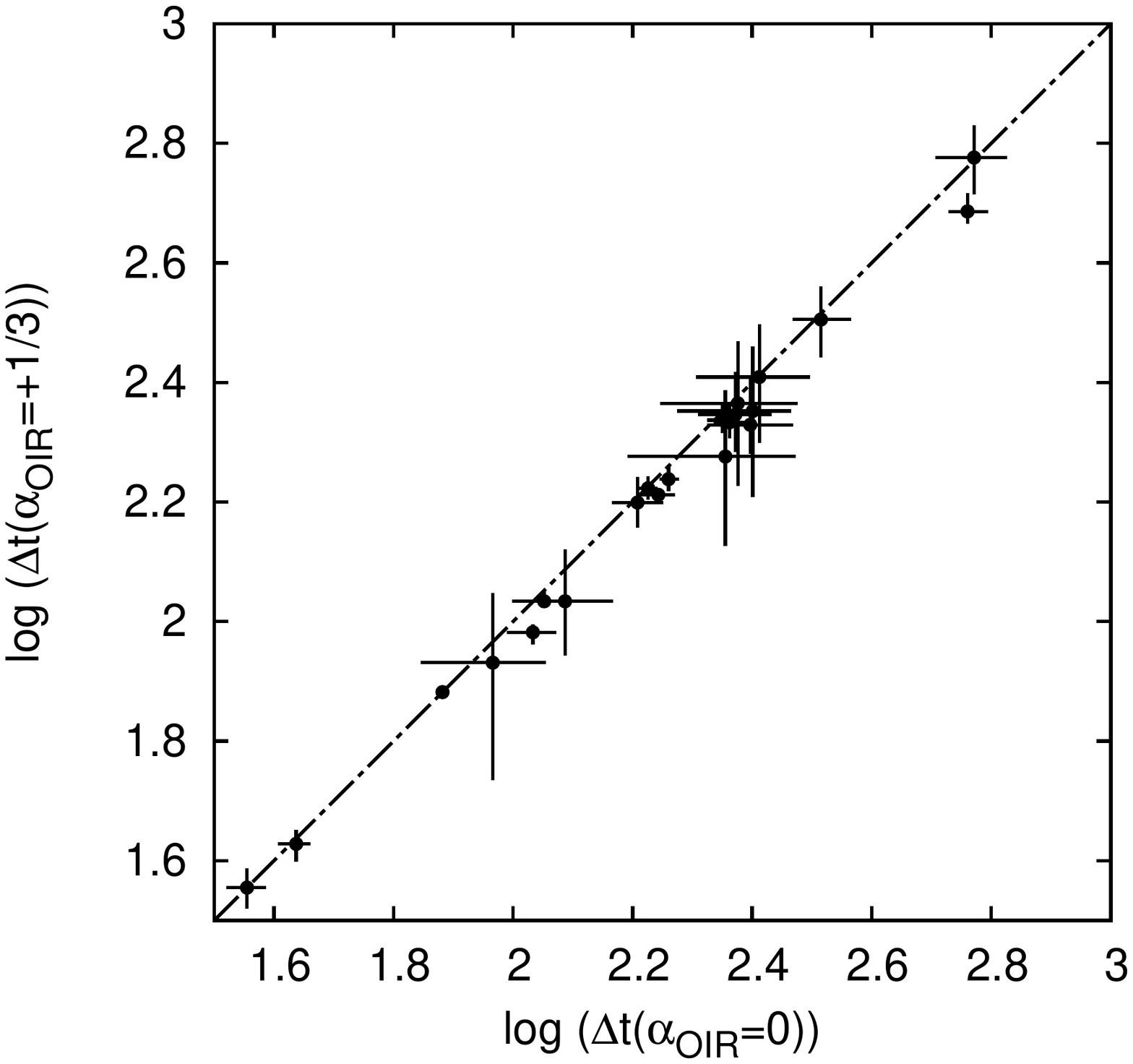}
 \caption{Comparison of the dust-reverberation lags assuming $\alpha_{\rm OIR}=0.0$ and $+1/3$ for the accretion-disk component in the optical and near-infrared wavelengths. The dotted-dashed line represents identical values.
 \label{fig:lagalpha}}
\end{figure}

\clearpage
\begin{figure}
\plotone{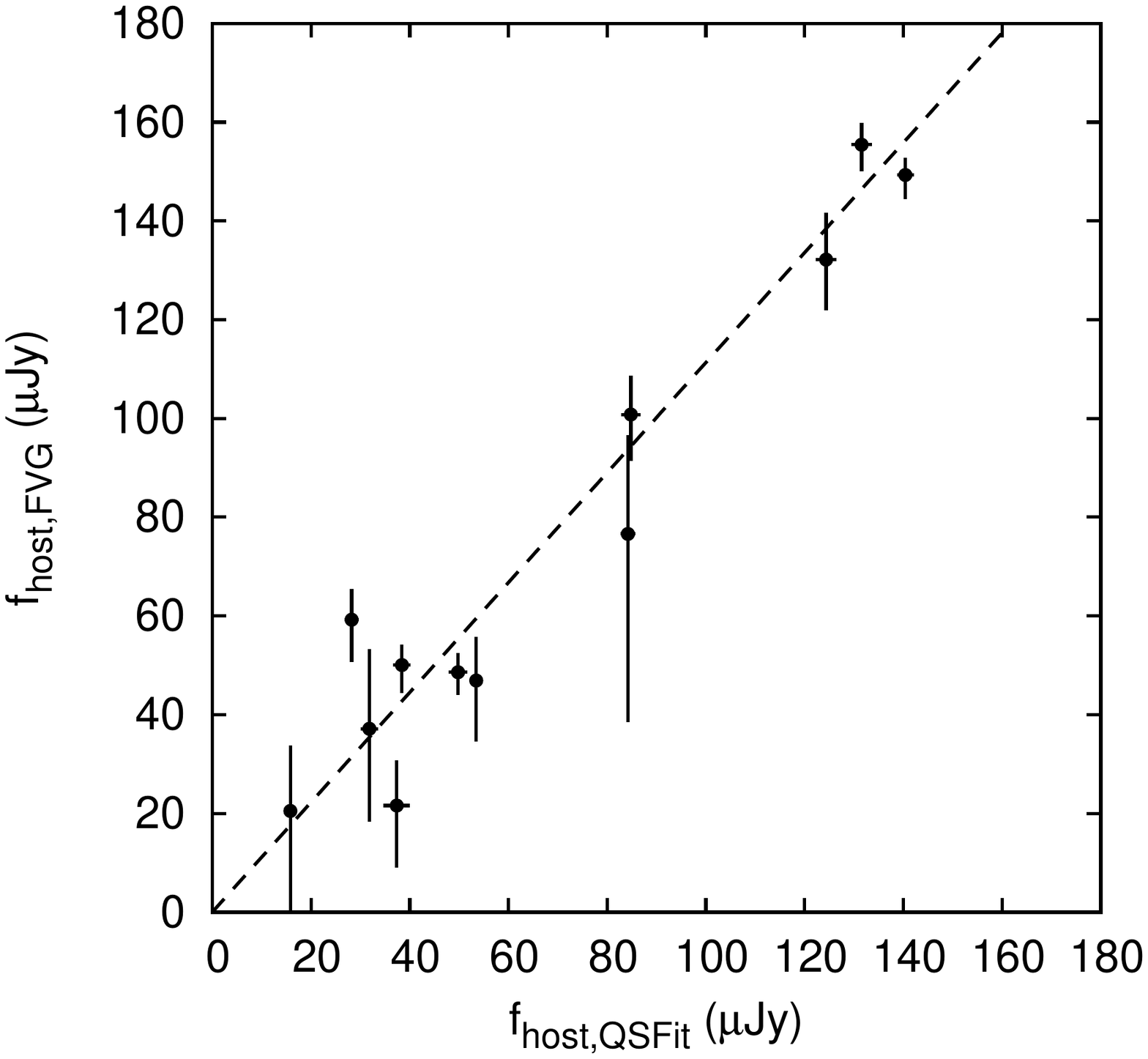}
 \caption{Comparison of the host-galaxy fluxes estimated by the spectral decomposition method ($f_{\rm host, QSFit}$) and that by the FVG method ($f_{\rm host, FVG}$). The dashed line represents the best-fit linear regression line.
 \label{fig:qsfit_fvg}}
\end{figure}

\clearpage
\begin{figure}
\plotone{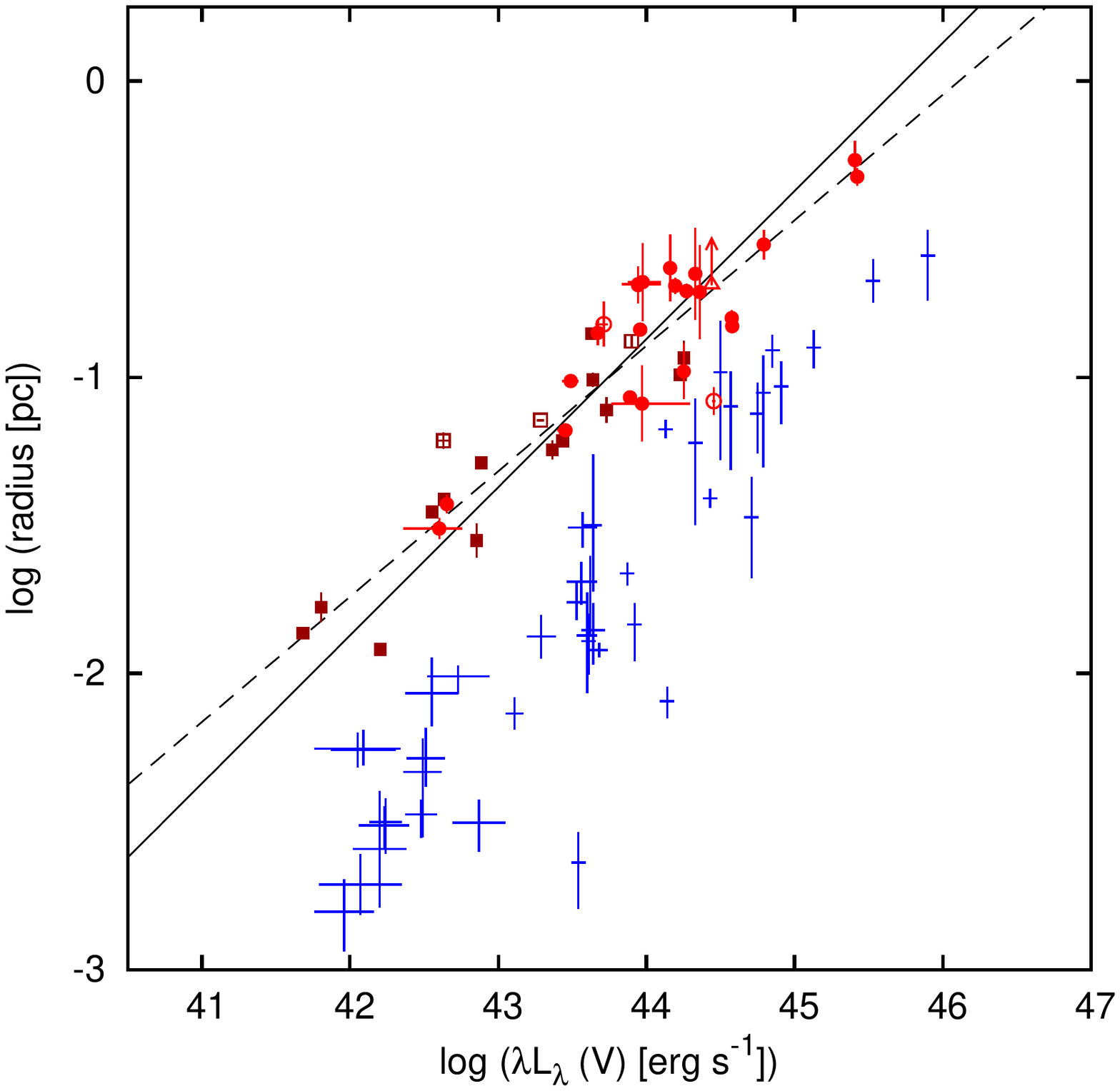}
 \caption{Radii of the innermost dust torus and the BLR plotted against the $V$-band luminosity. Red filled circles and brown filled boxes represent the dust-reverberation radii of this work and \citet{2014ApJ...788..159K}, respectively. Red open circles and the triangle represent the uncertain data and lower limit of the dust-reverberation radii of this work, and brown open boxes represent the data of \citet{2014ApJ...788..159K} for which the difference between the lag times obtained by JAVELIN and the CCF analysis was large. Blue crosses represent the reverberation radii of broad Balmer emission lines taken from \citet{2013ApJ...767..149B}. One data point per AGN is plotted, which was obtained by taking the median of the BLR lags, and the luminosity at that epoch, if more than one lag time was measured for an AGN. Solid and dashed lines represent the best-fit regression lines for the dust-reverberation radii with the slope $\beta $ being fixed at $0.5$ and freed, respectively. The data marked by the filled circles and boxes were used for the fitting. \label{fig:rLopt}}
\end{figure}

\clearpage
\begin{figure}
\plottwo{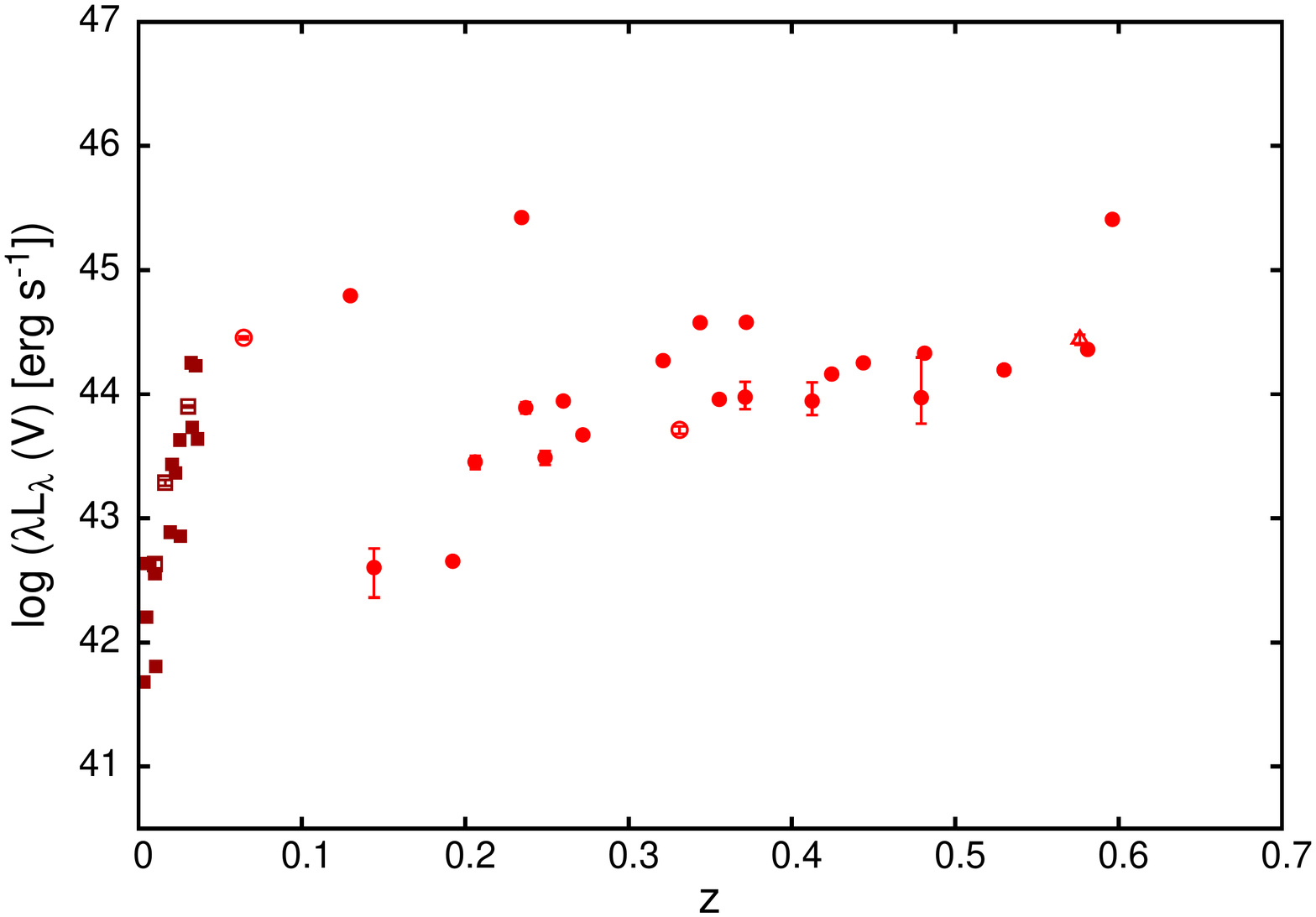}{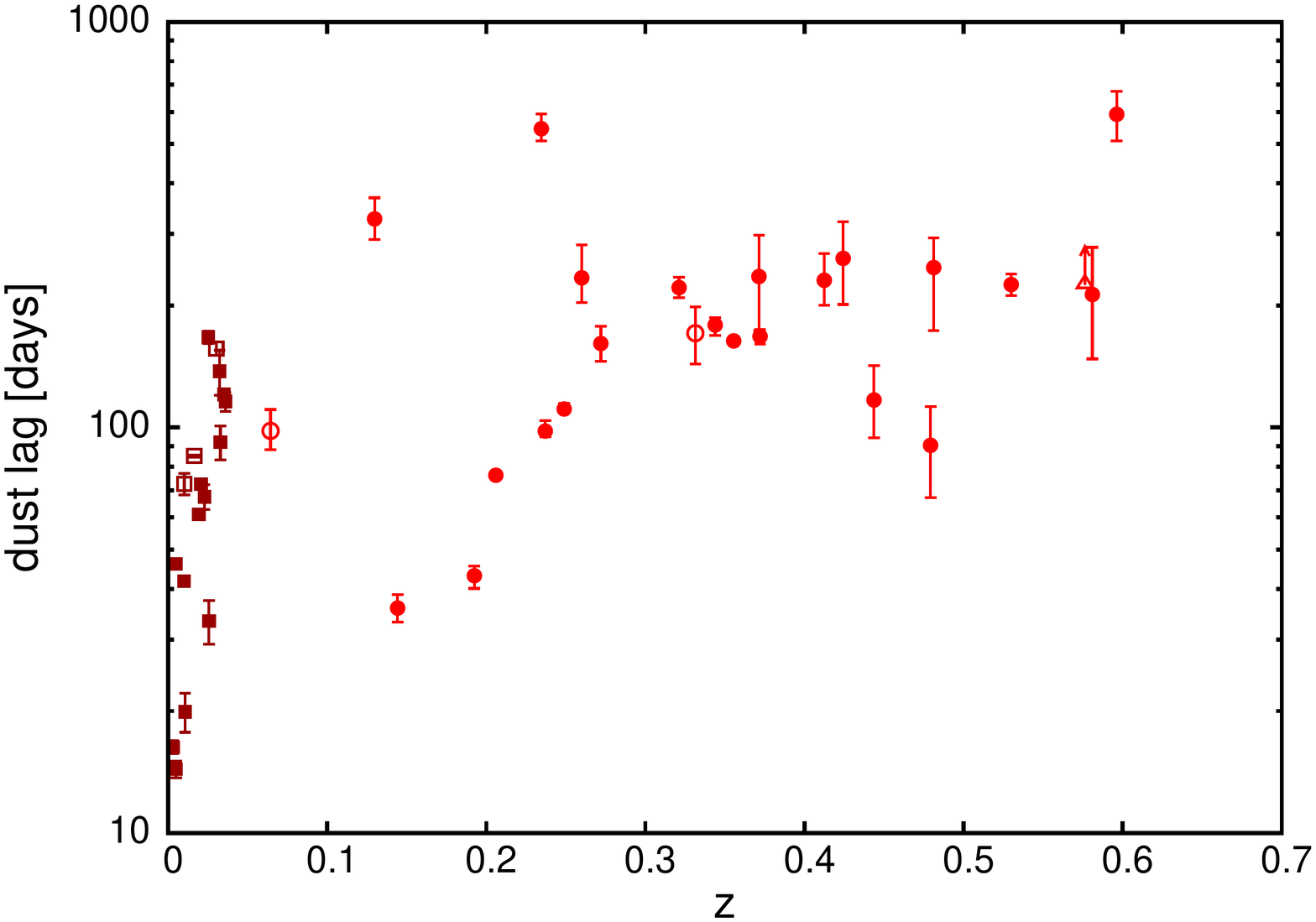}
 \caption{Luminosities (left panel) and the dust-reverberation lags (right panel) of the target AGNs plotted against their redshifts. Symbols are the same as Figure \ref{fig:rLopt}. The dust-reverberation lags are values in the observer frame.
 \label{fig:zllag}}
\end{figure}

\clearpage
\begin{figure}
\plotone{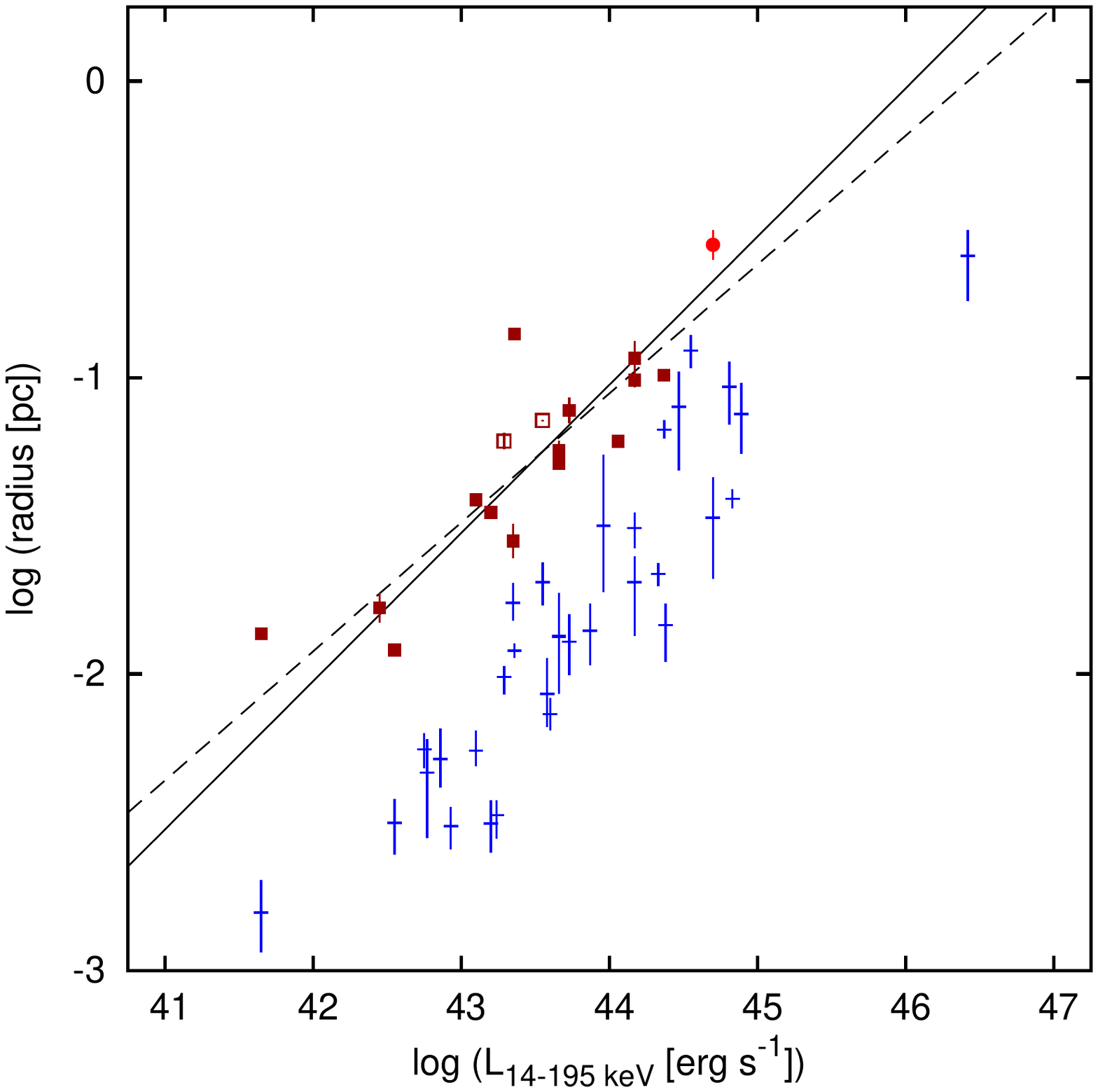}
 \caption{Radii of the innermost dust torus and the BLR plotted against the hard X-ray ($14-195$ keV) luminosity. Symbols are the same as in Figure \ref{fig:rLopt}. \label{fig:rLhx}}
\end{figure}

\clearpage
\begin{figure}
\plotone{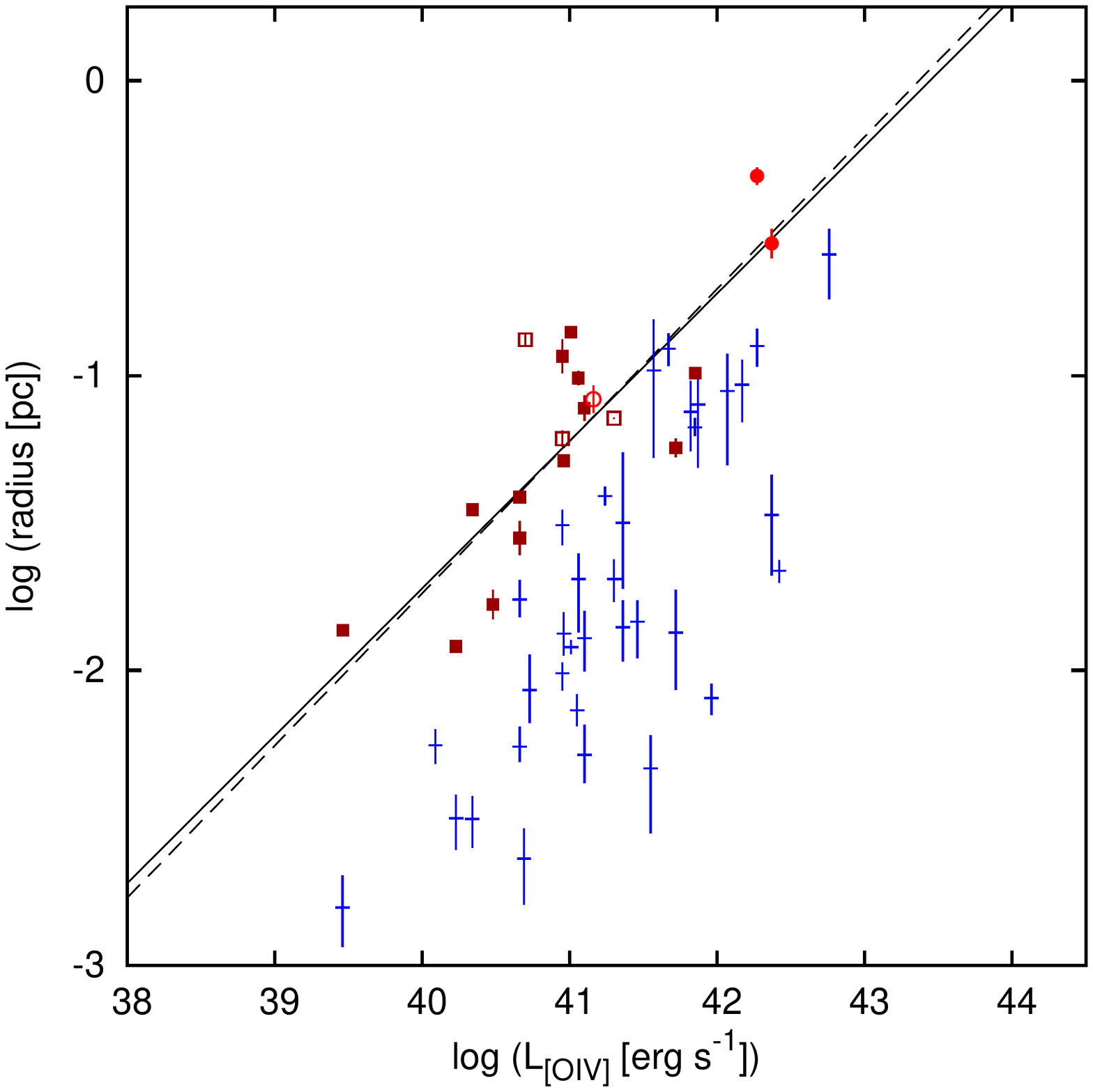}
 \caption{Radii of the innermost dust torus and the BLR plotted against the [OIV] ($26$ $\mu$m) emission-line luminosity. Symbols are the same as in Figure \ref{fig:rLopt}. \label{fig:rLoiv}}
\end{figure}

\clearpage
\begin{figure}
\plotone{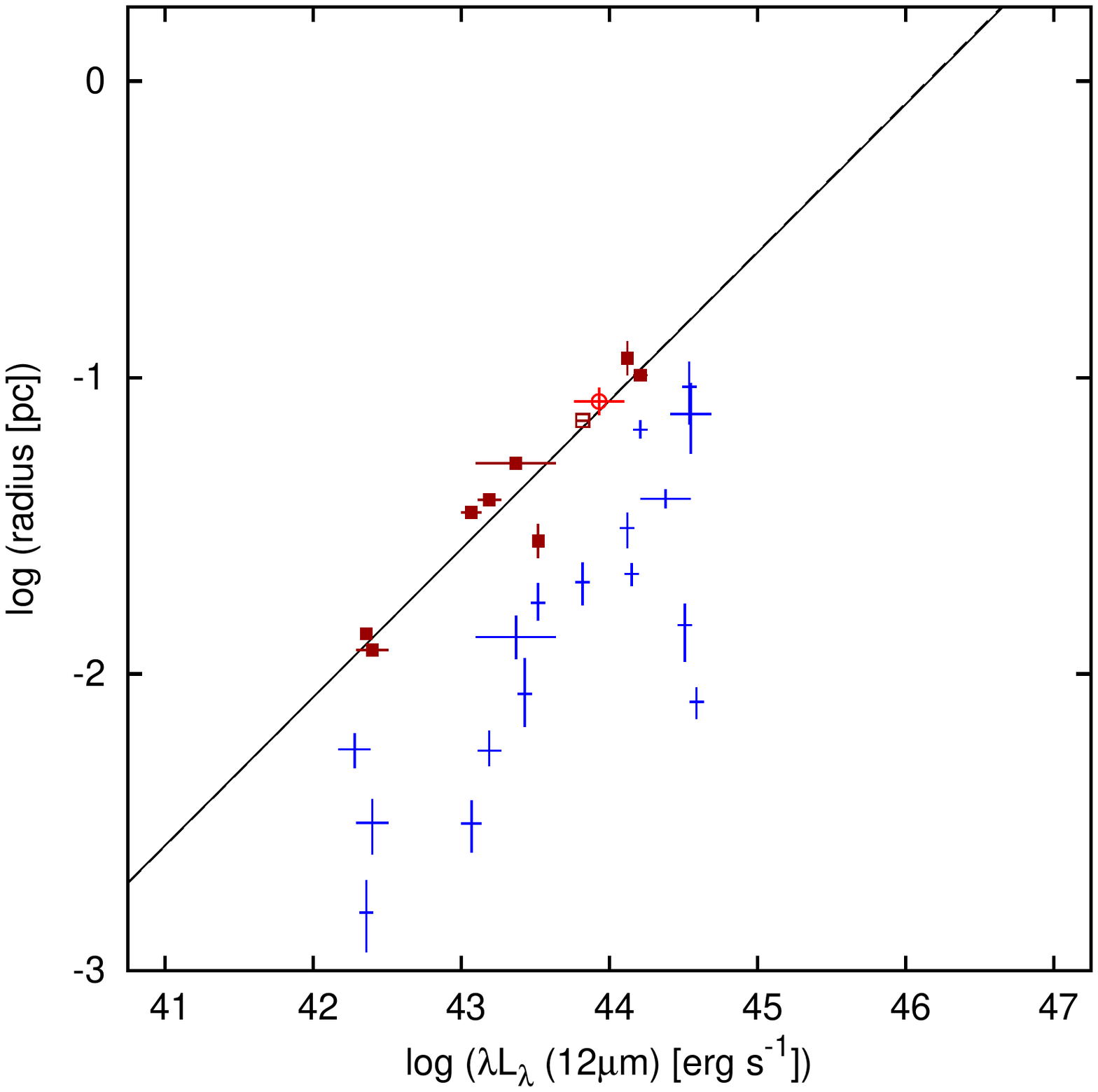}
 \caption{Radii of the innermost dust torus and the BLR plotted against the mid-infrared continuum emission ($\lambda $ $=12$ $\mu$m) luminosity. Symbols are the same as in Figure \ref{fig:rLopt}. \label{fig:rLmir}}
\end{figure}

\clearpage
\begin{figure}
\plotone{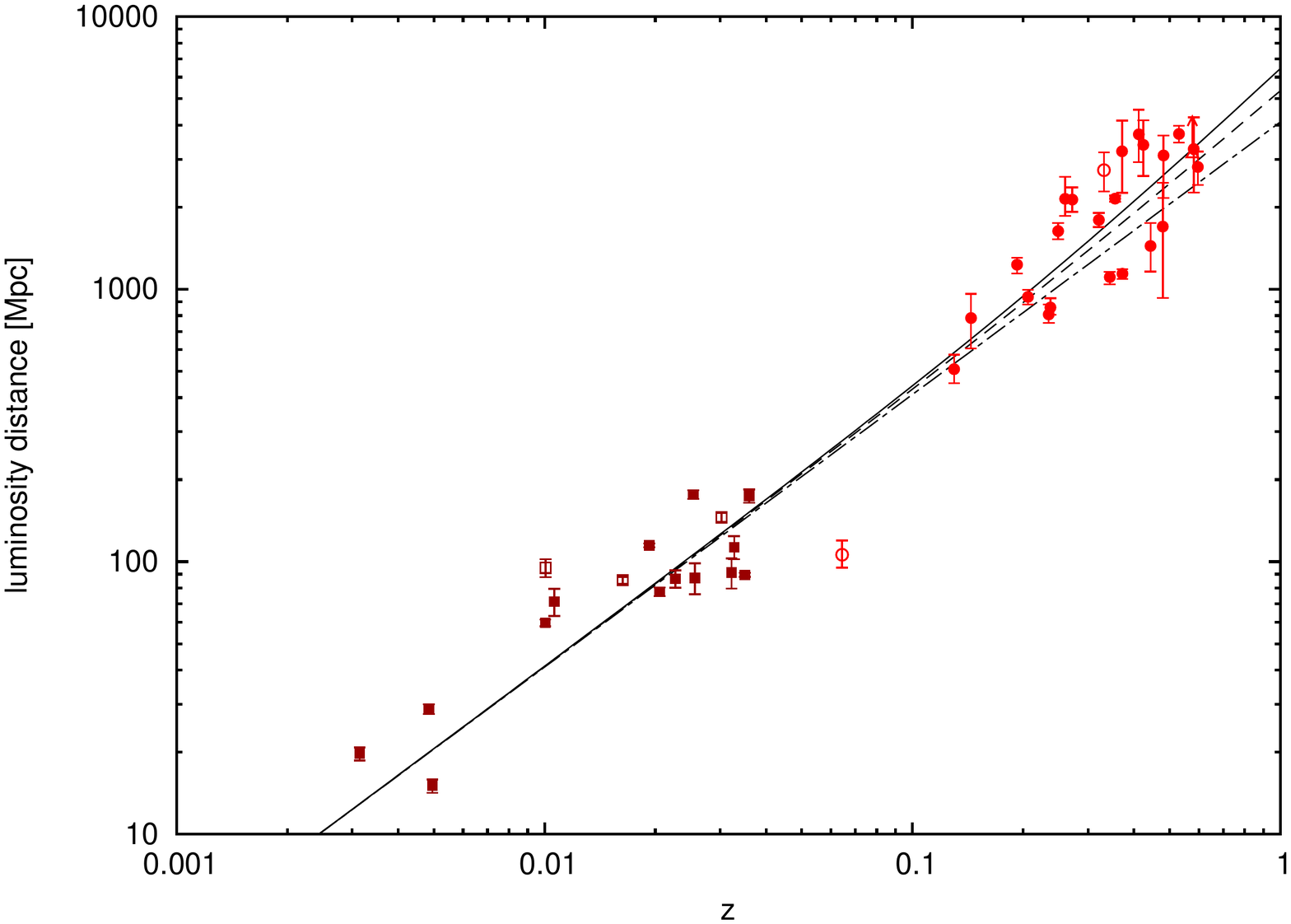}
 \caption{Diagram of luminosity distance based on the dust reverberation vs. redshift for the target AGNs. Symbols are the same as in Figure \ref{fig:rLopt}. The solid line represents the model curve for the cosmological parameters of $(\Omega _{0},\lambda _{0})=(0.27,0.73)$, the dashed line represents that of $q_0=0$, and the dotted-dashed line represents the relation $cz=H_0d$. The Hubble constant for all those lines is assumed to be $H_0=73$ km s$^{-1}$ Mpc$^{-1}$. \label{fig:hubble}}
\end{figure}

\end{document}